\documentclass[aps,preprint]{revtex4}
\usepackage{amsmath}
\usepackage[english]{babel}
\usepackage[latin1]{inputenc}
\usepackage{amssymb}
\usepackage{graphicx}
\usepackage{natbib}
\usepackage{epsfig}
\usepackage{bm}
\usepackage{dcolumn}

\begin{document}

\vspace*{2cm}

\title{\sc\Large{Strong magnetic fields in nonlocal chiral quark models}}

\author{D. G\'omez Dumm$^{a,b}$, M.F. Izzo Villafa\~ne$^{a,b}$, S. Noguera$^{c}$, V.P. Pagura$^{c}$ and N.N.\ Scoccola$^{b,d,e}$}

\affiliation{$^{a}$ IFLP, CONICET $-$ Departamento de F\'{\i}sica, Fac.\ de Cs.\ Exactas,
Universidad Nacional de La Plata, C.C. 67, (1900) La Plata, Argentina}
\affiliation{$^{b}$ CONICET, Rivadavia 1917, (1033) Buenos Aires, Argentina}
\affiliation{$^{c}$ Departamento de F\'{\i}sica Te\'{o}rica and IFIC, Centro Mixto
Universidad de Valencia-CSIC, E-46100 Burjassot (Valencia), Spain}
\affiliation{$^{d}$ Physics Department, Comisi\'{o}n Nacional de Energ\'{\i}a At\'{o}mica, }
\affiliation{Av.\ Libertador 8250, (1429) Buenos Aires, Argentina}
\affiliation{$^{e}$ Universidad Favaloro, Sol{\'{\i}}s 453, (1078) Buenos Aires, Argentina}

\begin{abstract}
We study the behavior of strongly interacting matter under a uniform intense
external magnetic field in the context of nonlocal extensions of the
Polyakov$-$Nambu$-$Jona-Lasinio model. A detailed description of the formalism
is presented, considering the cases of zero and finite temperature. In
particular, we analyze the effect of the magnetic field on the chiral
restoration and deconfinement transitions, which are found to occur at
approximately the same critical temperatures. Our results show that these
models offer a natural framework to account for the phenomenon of inverse
magnetic catalysis found in lattice QCD calculations.
\end{abstract}


\maketitle

\renewcommand{\thefootnote}{\arabic{footnote}}
\setcounter{footnote}{0}

\section{Introduction}

The study of the behavior of strongly interacting matter under intense
external magnetic fields has gained increasing interest in the last few
years. In fact, this topic has important applications e.g.~in the
description of compact objects like magnetars~\cite{duncan}, the analysis of
heavy ion collisions at very high energies~\cite{HIC} and the exploration of
the first phases of the Universe~\cite{cosmo}. Since these studies require
dealing with QCD in nonperturbative regimes,
present theoretical analyses are based either in the predictions of
effective models or in the results obtained through lattice QCD (LQCD)
calculations. In particular, the features of QCD phase transitions under
external magnetic fields deserve significant interest. Recent reviews on
this subject can be found in
Refs.~\cite{Kharzeev:2012ph,Andersen:2014xxa,Miransky:2015ava}. In view of
the difficulty of theoretical calculations, most works concentrate on the
case in which one has a uniform and static external magnetic field $\vec B$.
At zero temperature and chemical potential, both the results of low-energy
effective models of QCD and LQCD calculations indicate that the chiral quark
condensate should behave as an increasing function of $B$, which is usually
known as ``magnetic catalysis''. On the contrary, close to the chiral
restoration temperature, LQCD calculations carried out with realistic quark
masses~\cite{Bali:2011qj,Bali:2012zg} show that light quark-antiquark
condensates behave as nonmonotonic functions of the external magnetic field,
and this leads to a decrease of the transition temperature when the magnetic
field is increased. This effect is known as ``inverse magnetic catalysis''
(IMC). In addition, LQCD calculations predict an entanglement between the
chiral restoration and deconfinement critical
temperatures~\cite{Bali:2011qj}. These findings become a challenge to model
calculations. Indeed, most naive effective approaches to low-energy QCD
(Nambu$-$Jona-Lasinio model, chiral perturbation theory, MIT bag model,
quark-meson models) predict that the chiral transition temperature should
grow with $B$, i.e., they do not find IMC. In view of this discrepancy, in
the last few years, some more sophisticated low-energy effective models
compatible with the IMC effect have been proposed in the
literature~\cite{Skokov:2011ib,Fraga:2012ev,Bruckmann:2013oba,Bali:2013esa,Fukushima:2012kc,
Chao:2013qpa,Fraga:2013ova,Ferreira:2013tba,Ferreira:2014kpa,Ayala:2014iba,Farias:2014eca,
Ayala:2014gwa,Fayazbakhsh:2014mca,Andersen:2014oaa,Mueller:2015fka,Ayala:2014uua,Ferrer:2014qka,
Braun:2014fua,Ruggieri:2014bqa,Rougemont:2015oea,Ayala:2015bgv,Mao:2016fha}.
Possible mechanisms that allow the reproduction of IMC include, e.g., the
introduction of adequate ($B$-dependent) regularization prescriptions, or
explicit dependences of the effective coupling constants on the external
field. In particular, in the framework of the Nambu$-$Jona-Lasinio (NJL)
model, it has been shown that IMC can be obtained by considering a
$B$-dependent four-fermion coupling~\cite{Ayala:2014iba,Farias:2014eca}. On
the other hand, the problem of the entanglement between the deconfinement
and chiral restoration transitions has been studied in the context of the
Polyakov$-$Nambu$-$Jona-Lasinio (PNJL) model, in which fermions are coupled to a
background color field, and the traced Polyakov loop $\Phi$ is taken as
order parameter of the confinement/deconfinement transition. This extension
of the NJL model provides not only a description of confinement but also
allows one to obtain chiral restoration critical temperatures compatible with
those found in LQCD. In this framework, the effect of an external magnetic
field has been studied in Ref.~\cite{Gatto:2010pt}, where the authors
consider a Polyakov loop-dependent effective coupling constant in order to
avoid the splitting between chiral restoration and deconfinement
transitions. In this so-called ``entangled PNJL'' model, however, no IMC
effect is found (see also Refs.~\cite{Gatto:2012sp,Ferreira:2013tba}). Once
again, as shown in Ref.~\cite{Ferreira:2014kpa}, in the context of the PNJL
model one can reproduce lattice IMC results by considering a $B$-dependent
four-fermion coupling. Nevertheless, the results obtained in
Ref.~\cite{Ferreira:2014kpa} lead to a relatively large splitting ($\gtrsim
30$ MeV) between chiral restoration and deconfinement temperatures.

In this work we study the behavior of strongly interacting matter under a
uniform, static magnetic field in the framework of nonlocal chiral quark
models. This article is an extension of a previous work in which it has been
noticed that these kind of models offer a natural mechanism to understand
the IMC effect~\cite{Pagura:2016pwr}. Our aim is to present here a more
complete description of the formalism and also to extend the model to
incorporate the interaction with the Polyakov loop. As in the case of the
(local) NJL model, the traced Polyakov loop can be taken as an order
parameter of confinement, allowing one to describe simultaneously the chiral
restoration and deconfinement transitions. We will show that nonlocal models
are able to describe, at the mean field level, not only the IMC effect but
also the entanglement between both critical transition temperatures, in
quite reasonable agreement with LQCD results. The ``nonlocal PNJL'' (nlPNJL)
models considered here are a sort of nonlocal extensions of the PNJL model
that intend to provide a more realistic effective approach to QCD. In fact,
nonlocality arises naturally in the context of successful descriptions of
low-energy quark dynamics~\cite{Schafer:1996wv,RW94}, and it has been
shown~\cite{Noguera:2008} that nonlocal models can lead to a momentum
dependence in quark propagators that is consistent with LQCD results. It is
also found that in this framework one obtains an adequate description of the
properties of light mesons at both zero and finite
temperature/density~\cite{Noguera:2008,Bowler:1994ir,Schmidt:1994di,Golli:1998rf,
General:2000zx,Scarpettini:2003fj,GomezDumm:2004sr,GomezDumm:2006vz,
Contrera:2007wu,Hell:2008cc,Dumm:2010hh,Carlomagno:2013ona}. Moreover,
nlPNJL models (in the absence of interactions with external fields) provide a
description of the chiral restoration and deconfinement transitions that is
found to be in qualitative agreement with LQCD
calculations~\cite{Hell:2009by,Carlomagno:2013ona,Hell:2011ic,Kashiwa:2011td,Pagura:2011rt}.
As in Ref.~\cite{Pagura:2016pwr}, we consider here the case of nonlocal
quark models with separable interactions, using Ritus
eigenfunctions~\cite{Ritus:1978cj} to address the problem of including the
interaction with the magnetic field.

The article is organized as follows. In Sec.~II we start by introducing the
formalism to account for the presence of a constant magnetic field within
the framework of a nonlocal NJL-like model at zero temperature. Afterward,
we show how to extend this formalism to a finite temperature system, taking
also into account the coupling to the Polyakov loop. In Sec.~III we quote
our numerical results, discussing in detail the behavior of the different
relevant quantities as functions of the magnetic field and/or temperature.
In Sec.~IV we present our conclusions. Finally, in Appendixes A$-$D we give
some technical details concerning the derivation of various expressions
quoted in the main text.

\section{Theoretical formalism}

\subsection{Nonlocal NJL-like model in the presence of magnetic fields}

Let us start by stating the Euclidean action for our nonlocal NJL-like
two-flavor quark model,
\begin{equation}
S_E = \int d^4 x \ \left\{ \bar \psi (x) \left(- i \rlap/\partial
+ m_c \right) \psi (x) -
\frac{G}{2} j_a(x) j_a(x) \right\} \ .
\label{action}
\end{equation}
Here $m_c$ is the current quark mass, which is assumed to be equal for $u$
and $d$ quarks. The currents $j_a(x)$ are given by
\begin{eqnarray}
j_a (x) &=& \int d^4 z \  {\cal G}(z) \
\bar \psi(x+\frac{z}{2}) \ \Gamma_a \ \psi(x-\frac{z}{2}) \ ,
\label{cuOGE}
\end{eqnarray}
where $\Gamma_{a}=(\leavevmode\hbox{\small1\kern-3.8pt\normalsize1},i\gamma
_{5}\vec{\tau})$, and the function ${\cal G}(z)$ is a nonlocal form factor
that characterizes the effective interaction. We introduce now in the
effective action (\ref{action}) a coupling to an external
electromagnetic gauge field $\mathcal{A}_{\mu}$. 
For a local theory, this can be done by performing the replacement
\begin{equation}
\partial_{\mu}\ \rightarrow\ D_\mu\equiv\partial_{\mu}-i\,\hat Q
\mathcal{A}_{\mu}(x)\ ,
\label{covdev}
\end{equation}
where $\hat Q=\mbox{diag}(q_u,q_d)$, with $q_u=2e/3$, $q_d = -e/3$, is the
electromagnetic quark charge operator. In the case of the nonlocal model
under consideration, the inclusion of gauge interactions implies a change
not only in the kinetic terms of the Lagrangian but also in the nonlocal
currents in Eq.~(\ref{cuOGE}). One has
\begin{equation}
\psi(x-z/2) \ \rightarrow\ \mathcal{W}\left(  x,x-z/2\right)  \; \psi(x-z/2)\ ,
\label{transport}
\end{equation}
and a related change holds for $\bar
\psi(x+z/2)$~\cite{GomezDumm:2006vz,Noguera:2008,Dumm:2010hh}. Here, the
function $\mathcal{W}(s,t)$ is defined by
\begin{equation}
\mathcal{W}(s,t)\ =\ \mathrm{P}\;\exp\left[ -\, i \int_{s}^{t}dr_{\mu}\, 
\hat Q\mathcal{A}_{\mu}(r)
\right]  \ ,
\label{intpath}%
\end{equation}
where $r$ runs over an arbitrary path connecting $s$ with $t$. Regarding the
choice of this path, it is worth taking into account that none of the
procedures used to ``gauge'' theories that include nonlocal interactions
leads to a unique determination of the corresponding conserved
current~\cite{Mandelstam:1962mi}. The ambiguity, which in our case shows up
through the path choice for the line integral in Eq.~(\ref{intpath}), is
indeed present in any method used for the construction of a conserved
current from a nonlocal action. Its origin can be understood by noticing
that the condition of current conservation, which requires its divergence to
vanish, only fixes the longitudinal part of the current, the transverse part
remaining undetermined. This problem is well known in nuclear physics:
longitudinal components of exchange currents can be related to
phenomenological nucleon-nucleon forces, while transverse currents require a
specific model for the underlying meson exchanges~\cite{Gross:1987bu}.

Based on considerations of invariance and of simplicity, the straight line
path originally proposed in Ref.~\cite{Bloch:1952qkt} has been chosen
basically everywhere in the literature. Here we will also follow this choice,
parameterizing the path in Eq.~(\ref{intpath}) by
\begin{equation}
r_\mu = s_\mu + \lambda (t_\mu - s_\mu)\ ,
\label{stlp}
\end{equation}
with $\lambda$ running from 0 to 1. In the present context, this has to be
considered as a part of our model specification. In fact, although for some
particular observables the dependence on the path has been investigated and
found to be quite weak (see e.g.\ Refs.~\cite{Golli:1998rf,Dumm:2010hh}),
a thorough analysis of this issue is still lacking.

To proceed, it is convenient to bosonize the fermionic theory, introducing
scalar and pseudoscalar fields $\sigma(x)$ and $\vec{\pi}(x)$ and
integrating out the fermion fields. The bosonized action can be written
as~\cite{Noguera:2008,Dumm:2010hh}
\begin{equation}
S_{\mathrm{bos}}=-\ln\det\mathcal{D}_{x,x'}+\frac{1}{2G}
\int d^{4}x
\Big[\sigma(x)\sigma(x)+ \vec{\pi}(x)\cdot\vec{\pi}(x)\Big]\ ,
\label{sbos}
\end{equation}
with
\begin{eqnarray}
\mathcal{D}_{x,x'}   &  = & \delta^{(4)}(x-x')\,\big(-i\,\rlap/\!D + m_{c} \big)\,
+ \nonumber \\
& & \mathcal{G}(x-x') \, \gamma_{0} \, \mathcal{W} (x,\bar x)\,
\gamma_{0} \big[\sigma(\bar x) + i\,\gamma_5\,\vec{\tau}\cdot\vec{\pi}(\bar x) \big]
\, \mathcal{W}(\bar x,x') \ ,
\label{dxx}%
\end{eqnarray}
where $\bar x = (x+x')/2$ for the neutral mesons. We will consider the particular
case of a constant and homogenous magnetic field oriented along the
3-axis. To perform the analytical calculations we will use the Landau gauge,
in which one has $\mathcal{A}_\mu = B\, x_1\, \delta_{\mu 2}$. With this
gauge choice the function ${\cal W}(s,t)$ in Eq.~(\ref{intpath}) is given by
\begin{eqnarray}
{\cal W}(s,t) & = & \exp\left[-\frac{i}{2}\,\hat
Q\,B\,(s_1+t_1)\,(t_2-s_2) \right] \ .
\label{wstshwinger}
\end{eqnarray}
Next, we assume that the field $\sigma$ has a nontrivial translational
invariant mean field value $\bar{\sigma}$, while the mean field values of
pseudoscalar fields $\pi_{i}$ are zero. It should be stressed at this point
that the assumption stating that $\bar{\sigma}$ is independent of $x$ does
not imply that the resulting quark propagator will be translational
invariant. In fact, as discussed below, one can show that such an invariance
is broken by the appearance of the so-called Schwinger phase. Our assumption
just states that the deviations from translational invariance driven by the
magnetic field are not affected by the dynamics of the theory. In this way,
within the mean field approximation (MFA) we get
\begin{equation}
\mathcal{D}^{\mbox{\tiny MFA}}_{x,x'}=
 \ {\rm diag}
\big(
\mathcal{D}^{\mbox{\tiny MFA},u}_{x,x'} \, ,\,
\mathcal{D}^{\mbox{\tiny MFA},d}_{x,x'}
\big)\ ,
\label{diago}
\end{equation}
where
\begin{eqnarray}
\mathcal{D}^{\mbox{\tiny MFA},f}_{x,x'}  = \ \delta^{(4)}(x-x') \left( \Pi^f + m_c \right)
+ \, \bar\sigma \,\mathcal{G}(x-x') \,
\exp\left[i \Phi_f(x,x')\right]\ .
\label{df}
\end{eqnarray}
Here we have introduced the operator $\Pi^f = -i \rlap/\partial - q_f B \,
x_1 \gamma_2$, and a direct product to an identity matrix in color space is
understood. Notice that the second term on the rhs breaks translational
invariance through the Schwinger phase $\Phi_f(x,x')$, defined by
\begin{equation}
\Phi_f(x,x') \ \equiv \ (q_f B/2)\, (x_1+x'_1) \, (x_2-x'_2)\ ,
\label{schwingerphase}
\end{equation}
which arises from the product ${\cal W}(x,\bar x)\,{\cal W}(\bar x,x')$. In
this way, the MFA bosonized action per unit volume can be written as
\begin{eqnarray}
\frac{S^{\mbox{\tiny MFA}}_{\mathrm{bos}}}{V^{(4)}} & = & \frac{ \bar
\sigma^2}{2 G} - \frac{N_c}{V^{(4)}} \sum_{f=u,d}
\mbox{tr} \ln \mathcal{D}^{\mbox{\tiny MFA},f}_{x,x'}
\label{seff}
\end{eqnarray}
where in the second term of the rhs the traces over color and flavor have
been taken. To proceed to take the remaining traces over Dirac and
coordinate spaces it is convenient to perform the Ritus transform of
$\mathcal{D}^{\mbox{\tiny MFA},f}_{x,x'}$~\cite{Ritus:1978cj}. This is
defined by
\begin{equation}
\mathcal{D}^{\mbox{\tiny MFA},f}_{\bar p,\bar p\,'} = \int d^4x \ d^4x' \
\bar{\mathbb{E}}_{\bar p} (x)  \ \mathcal{D}^{\mbox{\tiny MFA},f}_{x,x'}  \
\mathbb{E}_{\bar p\,'} (x')\ ,
\label{dpp}
\end{equation}
where $\mathbb{E}_{\bar p} (x)$ and $\bar{\mathbb{E}}_{\bar p} (x)$, with
$\bar p=(k,p_2,p_3,p_4)$, are Ritus functions, the definitions and
properties of which are given in App.~A. The index $k$ is an integer that
will label the Landau energy levels. Using the properties of Ritus functions
we readily obtain
\begin{equation}
\mathcal{D}^{\mbox{\tiny MFA},f}_{\bar p,\bar p\,'}\ =\ \hat \delta_{\bar
p,\bar p\,'}\ P_{k,{s_f}} \left(- s_f \sqrt{2 k |q_f B|}\; \gamma_2 +
p_\parallel\cdot \gamma_\parallel + m_c \,\mathcal{I}\right) + \bar \sigma
\sum_{\lambda=\pm} G^{\lambda,f}_{\bar p,\bar p\,'}\, \Delta^\lambda\ ,
\end{equation}
where $\hat\delta_{\bar p,\bar p\,'}$ is a shorthand notation for
$(2\pi)^4\delta_{kk'}\,\delta(p_2-
p_2^{\;\prime})\,\delta(p_3-p_3^{\;\prime})\, \delta(p_4-p_4^{\;\prime})$,
and we have introduced the definitions $s_{f} = {\rm sign}(q_f B)$,
$p_\parallel = (p_3,p_4)$, $\gamma_\parallel = (\gamma_3,\gamma_4)$,
$\Delta^+=\mbox{diag}(1,0,1,0)$, $\Delta^-=\mbox{diag}(0,1,0,1)$ and
$P_{k,\pm 1}=(1-\delta_{k0})\,\mathcal{I}+\delta_{k0}\,\Delta^\pm$. The
functions $G^{\lambda,f}_{\bar p,\bar p\,'}$ are given by
\begin{equation}
G^{\lambda,f}_{\bar p,\bar p\,'} =  \int d^4x \ d^4x' \; E^\ast_{\bar
p\lambda} (x) \; \mathcal{G}(x-x') \, \exp\left[i \Phi_f(x,x')\right] \,
E_{\bar p\,'\lambda}(x')\ ,
\label{Gpp}
\end{equation}
the explicit form of $E_{\bar p\lambda}(x)$ being given in
Eq.~(\ref{autofuncion}). As is discussed in App. B, after some calculation
one can show that $G^{\lambda,f}_{\bar p,\bar p\,'}$ is, in fact, diagonal
in $\bar p,\bar p\,'$. One gets $G^{\lambda,f}_{\bar p,\bar p\,'} = \hat
\delta_{\bar p,\bar p\,'} \, g^{\lambda,f}_{k,p_\parallel}$, where
\begin{eqnarray}
g^{\lambda,f}_{k,p_\parallel} =
\frac{4\pi}{|q_fB|}\,(-1)^{k_\lambda}
\int \frac{d^2p_\perp}{(2\pi)^2}\
g(p_\perp^2 + p_\parallel^2) \,\exp(-p_\perp^2/|q_fB|) \,
L_{k_\lambda}(2p_\perp^2/|q_fB|)\ .
\label{mpk}
\end{eqnarray}
Here we have used the definitions $k_\pm = k - 1/2 \pm s_f/2\,$ and $p_\perp
= (p_1,p_2)$, while $g(p^2)$ is the Fourier transform of $\mathcal{G}(x)$
and $L_m(x)$ are Laguerre polynomials, with the usual convention $L_{-1}(x)
=0$. Defining now
\begin{equation}
M^{\lambda,f}_{k,p_\parallel} \ = \ \big(1-\delta_{k_\lambda,-1}\big)
m_c\, + \,\bar \sigma \, g^{\lambda,f}_{k,p_\parallel}\ ,
\label{mmain}
\end{equation}
we end up with $\mathcal{D}^{\mbox{\tiny MFA},f}_{\bar p,\bar p\,'} = \hat
\delta_{\bar p,\bar p\,'} \mathcal{D}^f_{k, p_\parallel}$, where
\begin{equation}
\mathcal{D}^f_{k, p_\parallel} \ = \ P_{k,{s_f}} \, \Big(\! -\! s_f\sqrt{2 k
|q_f B|}\; \gamma_2 +   p_\parallel \cdot \gamma_\parallel \Big) +
\sum_{\lambda=\pm} M^{\lambda,f}_{k,p_\parallel}\, \Delta^\lambda\ .
\label{twopoint}
\end{equation}

Then, using Eq.~(\ref{Opp}) and writing explicitly the trace over coordinate
space we have
\begin{equation}
\mbox{tr} \ln \mathcal{D}^{\mbox{\tiny MFA},f}_{x,x'} \ = \frac{N_c}{2\pi}\ \int d^4x\;
\sum_{k=0}^\infty
\frac{d^2p_\parallel}{(2\pi)^2}\; \int_{-\infty}^\infty \frac{dp_2}{2\pi}\, \mbox{tr}_{D} \Big[ \mathbb{E}_{\bar p}
(x)\; \ln \big(\mathcal{D}^f_{k,p_\parallel} \big)\, \mathbb{\bar E}_{\bar
p} (x) \Big] \ ,
\end{equation}
where $\mbox{tr}_D$ stands for the trace over Dirac space. Using the cyclic
property of the trace together with Eq.~(\ref{int2}), this expression reduces
to
\begin{equation}
\mbox{tr} \ln \mathcal{D}^{\mbox{\tiny MFA},f}_{x,x'} \ = \
V^{(4)} \, N_c\ \frac{|q_f B|}{2 \pi}
\sum_{k=0}^\infty \int \frac{d^2 p_\parallel}{(2 \pi)^2} \ \mbox{tr}_{D} \Big[
P_{k,{s_f}}\, \ln \big( \mathcal{D}^f_{k,p_\parallel} \big) \Big]\ .
\end{equation}
Since the matrix between the parentheses is not diagonal in Dirac space, it
is convenient to use at this stage the identity $\mbox{tr} \ln A = \ln \det
A$. After calculating the determinant and replacing in Eq.~(\ref{seff}), we
finally obtain
\begin{eqnarray}
\frac{S^{\mbox{\tiny MFA}}_{\mathrm{bos}}}{V^{(4)}} & = & \frac{ \bar
\sigma^2}{2 G} - N_c \sum_{f=u,d} \frac{  |q_f B|}{2 \pi} \int \frac{d^2p_\parallel}{(2\pi)^2}
\ \bigg[ \ln\left(p_\parallel^2 + {M^{\,\lambda_{\!
f},f}_{0,p_\parallel}\,}^2\,\right)
+  \sum_{k=1}^\infty \ \ln \Delta^f_{k,p_\parallel}\bigg]\ ,
\label{smfa}
\end{eqnarray}
where $\lambda_f = +\, (-)$ for $s_f = +1\,(-1)$, and
$\Delta^f_{k,p_\parallel}$ is defined by
\begin{equation}
\Delta^f_{k,p_\parallel} = \left( 2 k |q_f B| + p_\parallel^2 +
M^{+,f}_{k,p_\parallel}\, M^{-,f}_{k,p_\parallel} \right)^2\! +\, p_\parallel^2
\left( M^{+,f}_{k,p_\parallel} - M^{-,f}_{k,p_\parallel} \right)^2\ .
\end{equation}
Here, it is seen that the functions $M^{\pm,f}_{k,p_\parallel}$ play the role
of constituent quark masses in the presence of the external magnetic field.
The vacuum expectation value $\bar\sigma$ can now be found by minimizing the
effective action in Eq.~(\ref{smfa}). This leads to the gap equation
\begin{eqnarray}
\frac{\bar\sigma}{G} & = & N_c \sum_{f=u,d} \frac{|q_f B|}{\pi}
\sum_{k=0}^\infty \int \frac{d^2p_\parallel}{(2\pi)^2}
\sum_{\lambda=\pm}\,
\hat A^{\lambda,f}_{k,p_\parallel}\, g^{\lambda,f}_{k,p_\parallel}\ ,
\label{gapmain}
\end{eqnarray}
where we have defined
\begin{equation}
\hat A^{\pm,f}_{k,p_\parallel} \ = \ \frac{
M^{\mp,f}_{k,p_\parallel} \Big( 2 k |q_f B| + p_\parallel^2 +
M^{-,f}_{k,p_\parallel} M^{+,f}_{k,p_\parallel}\Big) + p_\parallel^2
\Big(M^{\pm,f}_{k,p_\parallel} - M^{\mp,f}_{k,p_\parallel}\Big)}
{\Delta^f_{k,p_\parallel}}\ \ .
\label{alam}
\end{equation}

Given the form of the two-point function in Eq.~(\ref{twopoint}), one can also
obtain the MFA quark propagators. Details of this calculation are given in
App.~C. In coordinate space, one gets
\begin{equation}
S^{\mbox{\tiny MFA},f}_{x,x'} \ = \ \big(\mathcal{D}^{\mbox{\tiny
MFA},f}_{x,x'}\big)^{-1} \ = \ \exp\!\big[i\Phi_f(x,x')\big]\,\int
\frac{d^4p}{(2\pi)^4}\ e^{i\, p\cdot (x-x')}\, \tilde
S^f(p_\perp,p_\parallel)\ ,
\end{equation}
where
\begin{eqnarray}
\tilde S^f(p_\perp,p_\parallel) & = & 2\, \exp(-p_\perp^2/|q_f B|)
\sum_{k=0}^\infty \sum_{\lambda=\pm} \Big[(-1)^{k_\lambda}
\big(\hat A^{\lambda,f}_{k,p_\parallel} - \hat B^{\lambda,f}_{k,p_\parallel}
\, p_\parallel\cdot\gamma_\parallel\big) L_{k_\lambda}(2p_\perp^2/|q_f B|)
+ \nonumber\\
& & 2\, (-1)^k \big(\hat C^{\lambda,f}_{k,p_\parallel}
- \hat D^{\lambda,f}_{k,p_\parallel}\, p_\parallel\cdot\gamma_\parallel\big)
\, p_\perp\cdot\gamma_\perp
\, L^1_{k-1}(2p_\perp^2/|q_f B|)\Big]\,\Delta^\lambda\ .
\label{sfp}
\end{eqnarray}
Here, we have introduced the definitions
\begin{eqnarray}
\hat B^{\pm,f}_{k,p_\parallel} & = & \hat C^{\pm,f}_{k,p_\parallel}
- M^{\mp,f}_{k,p_\parallel}\, \hat D^{\pm,f}_{k,p_\parallel}\ \ ,
\label{bb} \\
\hat C^{\pm,f}_{k,p_\parallel} & = & \frac{2 k |q_f B| + p_\parallel^2 +
M^{-,f}_{k,p_\parallel} M^{+,f}_{k,p_\parallel}}{\Delta^f_{k,p_\parallel}}
\ \ ,
\label{cc} \\
\hat D^{\pm,f}_{k,p_\parallel} & = & \frac{M^{\pm,f}_{k,p_\parallel}
- M^{\mp,f}_{k,p_\parallel}}{\Delta^f_{k,p_\parallel}} \ \ ,
\label{dd}
\end{eqnarray}
whereas $L_k^1(x)$ are generalized Laguerre polynomials, with $L^1_{-1} = 0$.
Notice that the functions $\hat A^{\lambda,f}_{k,p_\parallel}$ defined in
Eq.~(\ref{alam}) satisfy
\begin{equation}
\hat A^{\pm,f}_{k,p_\parallel} \ = \
M^{\mp,f}_{k,p_\parallel}\, \hat C^{\pm,f}_{k,p_\parallel} + p_\parallel^2
\, \hat D^{\pm,f}_{k,p_\parallel}\ .
\label{aa}
\end{equation}
As we have anticipated above, the quark propagators can be written as a
product of an exponential of the Schwinger phase times a translational
invariant function. It should be noticed that, as discussed in detail in
App.\ D, this form for the quark propagators (and the two-point functions)
is also obtained within the Schwinger-Dyson (SD) formalism using a general
ansatz as the one proposed in
Refs.~\cite{Leung:1996qy,Watson:2013ghq,Mueller:2014tea} [see
Eq.~(\ref{d6})]. Moreover, as shown in App.~D, in that framework one also
arrives at the gap equation quoted in Eq.~(\ref{gapmain}).

Given the quark propagators, the quark condensate for each flavor can be
easily calculated as
\begin{equation}
\langle\bar q_f \, q_f\rangle \ = \ -\,N_c \; \mbox{tr}_D \big[ S^{\mbox{\tiny
MFA},f}_{x,x} \big]\ .
\end{equation}
Alternatively, they can be obtained by taking the derivatives of
$S^{\mbox{\tiny MFA}}$ with respect to the current quark masses. The
associated explicit expressions, extended to the case of finite temperature,
will be given in the next subsection.

\subsection{Extension to finite temperature}

We extend now the analysis of the model introduced in the previous section
to a system at finite temperature. This is done by using the standard
Matsubara formalism. In order to account for confinement effects, we also
include the coupling of fermions to the Polyakov loop (PL), assuming that
quarks move on a constant color background field $\phi = i g\,\delta_{\mu
0}\, G^\mu_a \lambda^a/2$, where $G^\mu_a$ are the SU(3) color gauge fields.
We work in the so-called Polyakov gauge, in which the matrix $\phi$ is given
a diagonal representation $\phi = \phi_3 \lambda_3 + \phi_8 \lambda_8$,
taking the traced Polyakov loop $\Phi=\frac{1}{3} {\rm Tr}\, \exp( i
\phi/T)$ as an order parameter of the confinement/deconfinement transition.
Since---owing to the charge conjugation properties of the QCD
Lagrangian~\cite{Dumitru:2005ng}---the mean field traced Polyakov loop is
expected to be a real quantity, and $\phi_3$ and $\phi_8$ are assumed to be
real valued~\cite{Roessner:2006xn}, one has $\phi_8 = 0$, $\Phi = [1+ 2
\cos(\phi_3/T)]/3$. Finally, we include in the Lagrangian a term that
accounts for effective gauge field self-interactions, through a
Polyakov-loop potential ${\cal U}\,(\Phi, T)$. The resulting scheme is
usually denoted as nonlocal Polyakov$-$Nambu$-$Jona-Lasinio (nlPNJL)
model~\cite{Blaschke:2007np, Contrera:2007wu,
Contrera:2009hk,Hell:2008cc,Hell:2009by}.

Concerning the PL potential, its functional form is usually based on
properties of pure gauge QCD. In this work, we will mostly focus on a
potential given by a polynomial function based on a Ginzburg-Landau
ansatz~\cite{Ratti:2005jh,Scavenius:2002ru}, namely
\begin{eqnarray}
\frac{{\cal{U}}_{\rm poly}(\Phi ,T)}{T ^4} \ = \ -\,\frac{b_2(T)}{2}\, \Phi^2
-\,\frac{b_3}{3}\, \Phi^3 +\,\frac{b_4}{4}\, \Phi^4 \ ,
\label{upoly}
\end{eqnarray}
where
\begin{eqnarray}
b_2(T) = a_0 +a_1 \left(\dfrac{T_0}{T}\right) + a_2\left(\dfrac{T_0}{T}\right)^2
+ a_3\left(\dfrac{T_0}{T}\right)^3\ .
\label{pol}
\end{eqnarray}
The parameters $a_i$ and $b_i$  can be fitted to pure gauge lattice QCD
results imposing the presence of a first-order phase transition at $T_0$,
which is a further parameter of the model. In the absence of dynamical
quarks, from lattice calculations one expects a deconfinement temperature
$T_0 = 270$~MeV. However, it has been argued that in the presence of light
dynamical quarks this temperature scale should be adequately reduced to
about 210 and 190~MeV for the cases of two and three flavors, respectively,
with an uncertainty of about 30~MeV~\cite{Schaefer:2007pw}. The numerical
values for the parameters, taken from Ref.~\cite{Ratti:2005jh}, are
\begin{equation}
a_0 = 6.75\ ,\quad a_1 = -1.95\ ,\quad a_2 = 2.625\ ,\quad a_3 = -7.44
\ ,\quad b_3 = 0.75\ ,\quad b_4 = 7.5\ .
\end{equation}

It should be noticed that alternative forms for the PL potential have been
proposed in the literature. For example, an ansatz based on the logarithmic
expression of the Haar measure associated with the SU(3) color group
integration is considered in Ref.~\cite{Roessner:2006xn}, where its explicit
form and parameters can be found. Moreover, in Ref.~\cite{Haas:2013qwp} the
authors propose a so-called ``improved'' PL potential, in which the full QCD
potential ${\cal{U}}_{\rm glue}$ is related to that corresponding to the
pure Yang-Mills theory, ${\cal{U}}_{\rm YM}$, by
\begin{equation}
\frac{{\cal{U}}_{\rm glue}(\Phi ,t_{\rm glue})}{T ^4} \ = \
\frac{{\cal{U}}_{\rm YM}[\Phi ,t_{\rm YM}(t_{\rm glue})]}{T_{\rm YM}^4}\ ,
\label{utchica}
\end{equation}
where
\begin{equation}
t_{\rm YM}(t_{\rm glue}) \ = \ 0.57\, t_{\rm glue} \ = \
0.57 \left(\frac{T - T_c^{\rm glue}}{T_c^{\rm glue}}\right) \ .
\label{tglue}
\end{equation}
The dependence of the Yang-Mills potential on the Polyakov loop $\Phi$ and
the temperature $T_{YM}$ is taken from an ansatz such as that in
Eq.~(\ref{upoly}), while for $T_c^{\rm glue}$ a preferred value of 210~MeV
is obtained~\cite{Haas:2013qwp}. In our calculations we will also consider
these alternatives choices for the PL potential to get an estimation of the
possible qualitative impact on our results.

In this way, the grand canonical thermodynamic potential of the system under
the external magnetic field is found to be given by
\begin{eqnarray}
\Omega^{\mbox{\tiny MFA}}_{B,T} & = & \frac{ \bar
\sigma^2}{2 G} \ - \ T \sum_{n=-\infty}^{\infty} \sum_{c,f}\ \frac{  |q_f B|}{2 \pi} \int \frac{d p_3}{2\pi}
\ \bigg[ \ln\left({p_\parallel}_{nc}^2 + {M^{\lambda_{\!
f},f}_{0,{p_\parallel}_{nc}}}^{\!2}\, \right)
\ +
\nonumber \\
& & \ \sum_{k=1}^\infty \ \ln\left(
\Delta^f_{k,{p_\parallel}_{nc}}\right)\bigg] + \ {\cal U}(\Phi ,T)\ ,
\end{eqnarray}
where we have defined ${p_\parallel}_{nc} = (p_3\,,\,(2n+1)\pi T+\phi_c)$.
The sums over color and flavor indices run over $c=r,g,b$ and $f=u,d$,
respectively, while the color background fields are $\phi_r = - \phi_g =
\phi_3$, $\phi_b = 0$. As usual in nonlocal models, it is seen that
$\Omega^{\mbox{\tiny MFA}}$ turns out to be divergent, and thus it has to be
regularized. We use a prescription similar to that considered, e.g., in
Ref.~\cite{GomezDumm:2004sr}, namely
\begin{equation}
\Omega^{{\mbox{\tiny MFA}},\rm reg}_{B,T}\ = \ \Omega^{\mbox{\tiny
MFA}}_{B,T}\, -\, \Omega^{\rm free}_{B,T}\, +\, \Omega^{\rm free,reg}_{B,T}\ .
\end{equation}
Notice that here the ``free'' potential keeps the interaction with the
magnetic field and the PL; i.e., only $\bar \sigma$ is set to zero. For this
``free'' piece the Matsubara sum can be performed analytically, leading to
\begin{eqnarray}
\Omega^{\rm free,reg}_{B,T}
&=& -\ \frac{N_c}{2\pi^2} \, \sum_f\ (q_f B)^2
\left[ \zeta'(-1,x_f) + \frac{x_f^2}{4} - \frac{1}{2} (x_f^2 -x_f)\ln x_f
\right]\; -
\nonumber \\
& & T \sum_{f,c}\;
\frac{|q_f B|}{\pi}  \sum_{k=0}^\infty \alpha_k
\int \frac{d p}{2\pi}\; \ln \bigg\{ 1 + \exp\left[- (\epsilon^f_{kp} + i \phi_c)/T\right]
\bigg\} \ ,
\end{eqnarray}
where $x_f = m_c^2/(2 |q_f B|)$, $\alpha_k = 2-\delta_{k0}$, and
$\epsilon^f_{kp} = (2 k |q_f B| + p^2 + m_c^2)^{1/2}$. In addition,
$\zeta'(-1,x_f) = d \zeta(z,x_f)/dz |_{z=-1}$, where $\zeta(z,x_f)$ is the
Hurwitz zeta function. Owing to the presence of the background field, one
has now a set of two coupled ``gap equations''
\begin{equation}
\frac{\partial{\Omega^{{\mbox{\tiny MFA}},\rm reg}_{B,T}} }{\partial \bar
\sigma} \ = \ 0\ , \qquad
\frac{\partial{\Omega^{{\mbox{\tiny MFA}},\rm reg}_{B,T}
}}{\partial \Phi} = 0\ .
\end{equation}

Given $\Omega^{\mbox{\tiny
MFA},reg}_{B,T}$, the magnetic field-dependent quark condensate for each flavor can be
calculated by taking the derivative with respect to the corresponding
current quark mass. This leads to
\begin{eqnarray}
\langle \bar q_f q_f\rangle^{\rm reg}_{B,T} &=& -\ \frac{|q_f B|\, T}{\pi}
\sum_c \int \frac{d p_3}{2\pi} \sum_{k=0}^\infty \sum_{n=-\infty}^\infty
\left(\,\sum_{\lambda=\pm} \hat A^{\lambda,f}_{k,{p_\parallel}_{nc}}
 - \frac{ 2 m_c}{p_{\parallel_{nc}}^{\,2} + 2 k |q_f B| + m_c^2} \right) -
 \nonumber \\
&& \frac{N_c m_c^3}{4 \pi^2}
\left[ \frac{\ln \Gamma(x_f)}{x_f} - \frac{\ln 2 \pi}{2 x_f} + 1 - \left(1-\frac{1}{2x_f}\right) \ln x_f
\right] +
\nonumber \\
&& \frac{|q_f B|}{\pi} \sum_c \sum_{k=0}^\infty \alpha_k
\int \frac{d p}{2\pi}\,\ \frac{m_c}{\epsilon^f_{kp}}\, \
\frac{1}{1 +
\exp[(\epsilon^f_{kp} + i \phi_c)/T]}\;\ .
\label{condensate}
\end{eqnarray}

Finally, to make
contact with the LQCD results quoted in Ref.~\cite{Bali:2012zg} we define
the quantity
\begin{equation}
\Sigma^f_{B,T} \ = \ -\frac{2\, m_c}{S^4} \left[ \langle \bar q_f q_f \rangle^{\rm reg}_{B,T}
 - \langle \bar q q\rangle^{\rm reg}_{0,0} \right] + 1 \ ,
\label{defi}
\end{equation}
where $S$ is a phenomenological scale fixed as $S= (135 \times 86)^{1/2}$
MeV. The subindex $f$ can be omitted for $B = 0$, owing to isospin symmetry.
We also introduce the definitions $\Delta \Sigma^f_{B,T} = \Sigma^f_{B,T} -
\Sigma^f_{0,T}$, $\bar \Sigma_{B,T} = (\Sigma^u_{B,T}+\Sigma^d_{B,T})/2$ and
$\Delta \bar \Sigma_{B,T} = (\Delta \Sigma^u_{B,T}+\Delta
\Sigma^d_{B,T})/2\,$, which correspond to the subtracted normalized flavor
condensate, the normalized flavor average condensate, and the subtracted
normalized flavor average condensate, respectively.

\section{Numerical results}

To obtain numerical predictions for the behavior of the above-defined
quantities as functions of the temperature and the external magnetic field,
it is necessary to specify the particular shape of the nonlocal form factor
$g(p^2)$. We consider here two often-used forms, namely a Gaussian
function,
\begin{equation}
g(p^2) \ = \ \exp(-p^2/\Lambda^2)
\end{equation}
and a ``5-Lorentzian'' function,
\begin{equation}
g(p^2) \ = \ \frac{1}{1 + (p^2/\Lambda^2)^5}\ \ .
\end{equation}
Notice that in these form factors we introduce an energy scale $\Lambda$,
which acts as an effective momentum cutoff. This has to be taken as an
additional parameter of the model. The functions $g(p^2)$ are normalized to
$g(0) = 1$, which is equivalent to the condition \mbox{$\int d^4z\;{\cal
G}(z) = 1$} for the form factors in coordinate space. In any case, this
condition can be relaxed by redefining the coupling constant $G$ in the
Lagrangian. In the particular case of the Gaussian function, one has the
advantage that the integral in Eq.~(\ref{mpk}) can be performed
analytically. One gets
\begin{equation}
M^{\lambda,f}_{\bar p,k} \ = \ \big(1-\delta_{k_\lambda,-1}\big) m_c \ + \
\bar \sigma \; \frac{ \left(1- |q_f B|/\Lambda^2\right)^{k_\lambda}}
{\left(1+ |q_f B|/\Lambda^2\right)^{k_\lambda+1}}
\;\exp\!\big(-{\bar p}^{\,2}/\Lambda^2\big)\ .
\label{gauss}
\end{equation}

Given the nonlocal form factor, one has to determine the values of the
parameters $m_c$, $G$ and $\Lambda$. Here, we will consider different
parameter sets, obtained by requiring that the model leads to the empirical
values of the pion mass and decay constant, as well as some chosen value of
the quark condensate $\langle \bar q q\rangle^{\rm reg}_{0,0}$. We will
consider in particular the phenomenologically acceptable values $(-\langle
\bar q q\rangle^{\rm reg}_{0,0})^{1/3} = 220$, 230, and 240~MeV. The
corresponding parameter sets for the Gaussian and 5-Lorentzian form factors
are quoted in Table~\ref{tab1}. The analytical expressions used to calculate
the values of the pion mass and decay constant within the nonlocal NJL model
can be found, e.g., in Ref.~\cite{GomezDumm:2006vz}.

\begin{table}
\caption{Model parameters for Gaussian and 5-Lorentzian form factors leading
to some representative values of the chiral condensate}
\label{tab1}
\begin{center}
\begin{tabular}{ccccc}
\ $(-\langle\, q \bar q\,\rangle_{0,0}^{\rm reg})^{1/3}$ (MeV) \ & \ Form factor \ &
\ \ \ $m_c$ (MeV) \ \ \ & \ \  $G \Lambda^2$  \ \  & \ \ \ \ $\Lambda$ (MeV) \ \ \ \ \\
\hline
220                  &     G        &  7.4     &   29.06        &    604     \\
\cline{1-5}
                     &     L5       &  7.4     &   10.34        &    790     \\
\hline
230                  &     G        &  6.5     &   23.66        &    678     \\
\cline{1-5}
                     &     L5       &  6.5     &   9.700        &    857     \\
\hline
240                  &     G        &  5.8     &   20.65        &    752     \\
\cline{1-5}
                     &     L5       &  5.8     &   9.267        &    926     \\
\hline
\end{tabular}
\end{center}

\end{table}

\begin{figure}[hbt]
\includegraphics[width=0.95\textwidth]{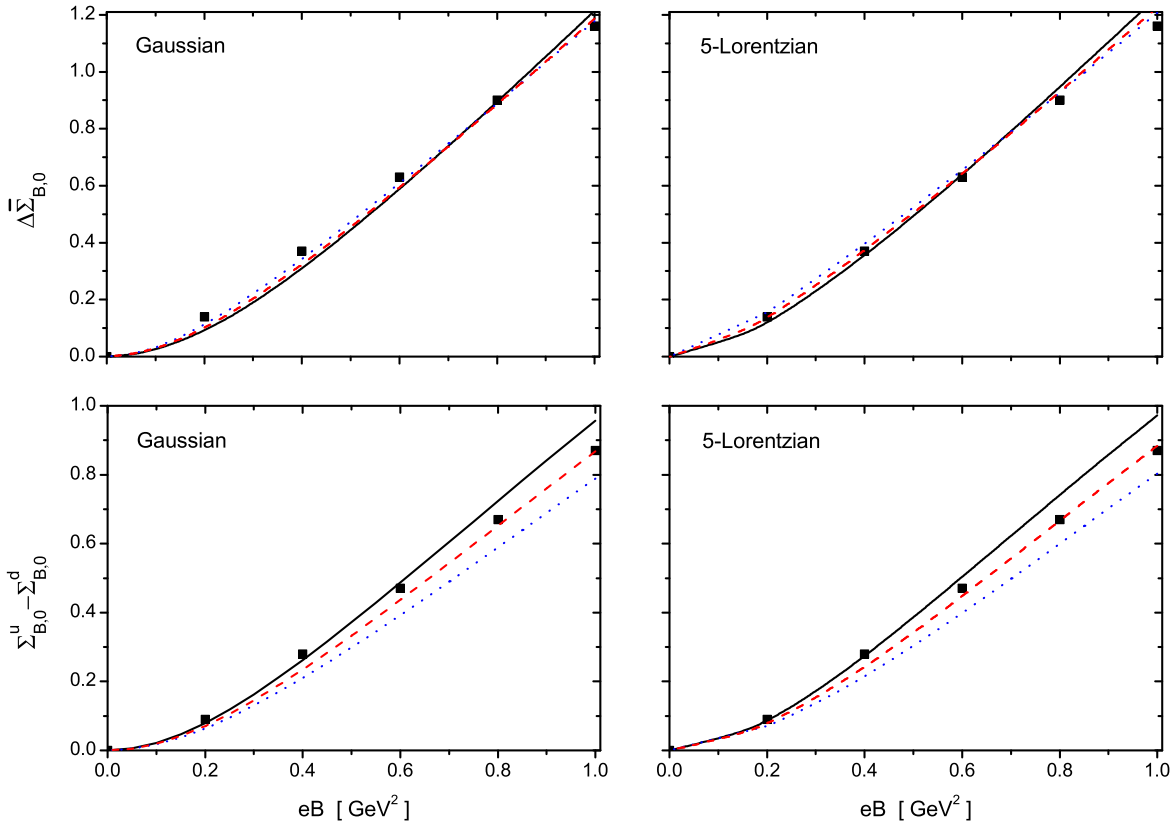}
\caption{Normalized condensates as functions of the magnetic field at $T =
0$. Upper panel: subtracted flavor average; lower panel: flavor difference
[see Eq.~(\ref{defi}) and the text below]. Solid (black), dashed (red), and
dotted (blue) curves correspond to parameterizations leading to $(-
\langle\bar q q \rangle^{\rm reg}_{0,0})^{1/3} = 220$, 230, and 240 MeV,
respectively. Full square symbols indicate LQCD results taken from
Ref.~\cite{Bali:2012zg}. } \label{fig1}
\end{figure}

Let us start by discussing our results for zero temperature. In the upper
panels of Fig.~\ref{fig1} we show the model predictions for $\Delta \bar
\Sigma_{B,0}$ as a function of $eB$ for various model parametrizations,
while in the lower panels we show the corresponding results for
$\Sigma^u_{B,0}-\Sigma^d_{B,0}$. LQCD data from Ref.~\cite{Bali:2012zg} are
also displayed in both cases for comparison. Solid, dashed, and dotted
curves correspond to $(-\langle \bar q_f q_f\rangle^{\rm reg}_{0,0})^{1/3} =
220$, 230, and 240 MeV, respectively. It can be seen that the predictions
for $\Delta \bar \Sigma_{B,0}$ are very similar for all considered
parametrizations, showing a very good agreement with LQCD results. In the
case of $\Sigma^u_{B,0}-\Sigma^d_{B,0}$, although the overall agreement with
LQCD calculations is still good, we find some dependence on the
parameterization. As shown in the figure, for both form factor shapes the
parameter sets leading to a condensate of $(-\langle\bar q_f q_f
\rangle^{\rm reg}_{0,0})^{1/3} = 230$ MeV seem to be preferred.

\begin{figure}[hbt]
\includegraphics[width=0.75\textwidth]{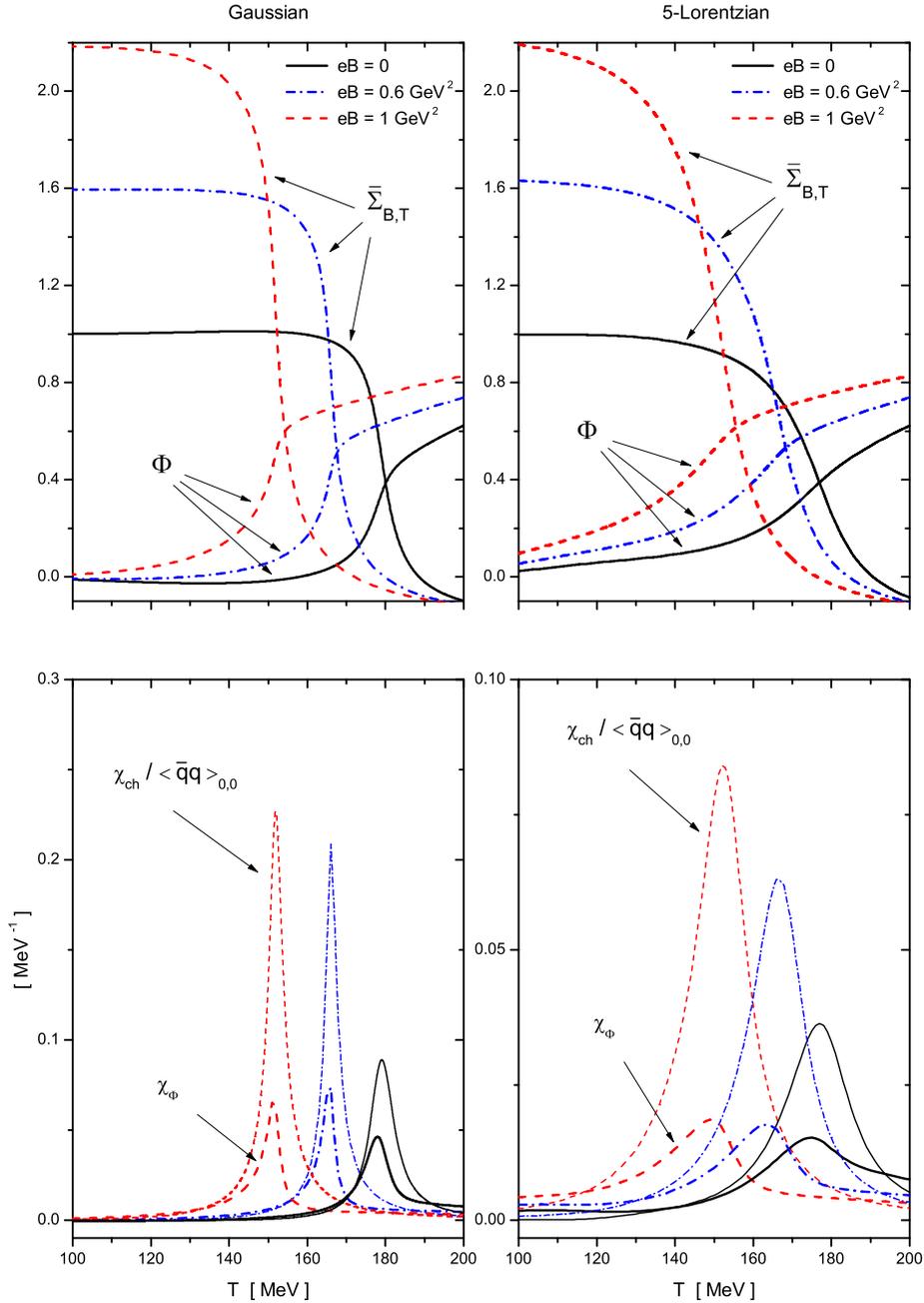}
\caption{Upper panels: normalized flavor average condensate and traced
Polyakov loop as functions of the temperature, for three representative
values of $eB$. Lower panels: behavior of the corresponding chiral and PL
susceptibilities as functions of the temperature.} \label{fig2}
\end{figure}

We turn now to our numerical results for a system at finite temperature. In
the upper panels of Fig.~\ref{fig2} we show the behavior of the averaged
chiral condensate $\bar\Sigma_{B,T}$ and the traced Polyakov loop $\Phi$ as
functions of the temperature, for three representative values of the
external magnetic field $B$, namely $B = 0$, 0.6, and 1 GeV$^2$. The curves
correspond to parameter sets leading to $(-\langle\bar q q \rangle^{\rm
reg}_{0,0})^{1/3} = 230$ MeV and a polynomial Polyakov-loop potential with
$T_0 = 210$ MeV. Given a value of $B$, it is seen from the figure that for
the cases of both Gaussian and 5-Lorentzian form factors the chiral
restoration and deconfinement transitions proceed as smooth crossovers, at
approximately the same critical temperatures. For definiteness, we take these
temperatures from the maxima of the chiral and PL susceptibilities, which we
define as the derivatives $\chi_{\rm ch} = - \partial[(\langle\bar uu
\rangle^{\rm reg}_{B,T} + \langle\bar dd \rangle^{\rm reg}_{B,T})/2]/\partial
T$ and $\chi_\Phi = \partial\Phi/\partial T$, respectively. Our results for
the behavior of the susceptibilities as functions of the temperature, for $B
=0$, 0.6, and 1 GeV$^2$, are shown in the lower panels of Fig.~\ref{fig2}.

\begin{table}
\caption{Critical temperatures for $B=0$ and various parametrizations.}
\label{tab2}
\begin{center}
\begin{tabular}{l|ccc|ccc}
& \multicolumn{3}{c}{\ \ Gaussian\ \ }  & \multicolumn{3}{c}{5-Lorentzian} \\
$(-\langle\, q \bar q\,\rangle_{0,0}^{\rm reg})^{1/3}$ (MeV) \ & \ 220 & 230 &
240 \ & \ 220 & 230 & 240 \ \\
\hline
Chiral $T_c$ (MeV) & \ \ 182.1 \ & \ 179.1 \ & \ 177.4 \ \ &
\ \ 177.0 \ & \ 177.0 \ & \ 177.8 \ \ \\
Deconfinement $T_c$ (MeV)\ \ & \ \ 182.1 \ & \ 178.0 \ & \ 175.8 \ \ &
\ \ 174.8 \ & \ 174.7 \ & \ 175.5 \ \
\end{tabular}
\end{center}

\end{table}

The chiral restoration and deconfinement critical temperatures obtained in
the absence of external magnetic field for different parametrizations are
quoted in Table~\ref{tab2}. It is seen that in all cases the splitting
between both critical temperatures is below 5 MeV, which is consistent with
the results obtained in lattice QCD. From Table~\ref{tab2} it is also seen
that the values of critical temperatures do not vary significantly with the
parametrization (recalling that in all cases the parameters have been fixed
to reproduce the empirical values of the pion mass and decay constant). On
the other hand, the critical temperatures in Table~\ref{tab2} are found to
be somewhat higher than those obtained from LQCD, which lie around
160~MeV~\cite{Aoki:2009sc,Bazavov:2011nk}. In fact, the value of $T_c$ and
the steepness of the transition depend on the form of the Polyakov-loop
potential. It is found that the logarithmic PL
potential~\cite{Roessner:2006xn} leads in general to steep transitions
(which can be even of first order for certain values of the parameters),
whereas the ``improved'' PL potentials [see Eqs.~(\ref{utchica}) and
(\ref{tglue})] lead to a smoother behavior that shows a better agreement
with LQCD results~\cite{Carlomagno:2013ona}. In particular, for an
``improved polynomial'' PL potential, one can get $T_c\simeq 160$ to 165 MeV,
depending on the parametrization. It is worth noticing that in the absence of
the interaction with the Polyakov loop the values of $T_c$ drop down to
about 130~MeV~\cite{Pagura:2016pwr}.

\begin{figure}[hbt]
\includegraphics[width=0.8\textwidth]{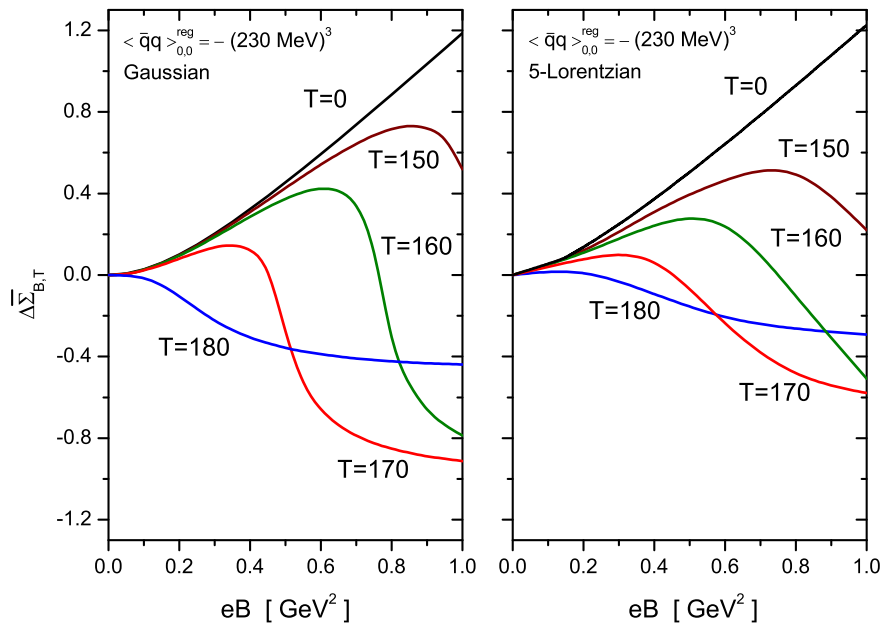}
\caption{Subtracted normalized flavor average condensate as a function of
$eB$ for different representative temperatures. Left and right panels
correspond to Gaussian and 5-Lorentzian form factors, respectively, with
$(-\langle\bar q q \rangle^{\rm reg}_{0,0})^{1/3} = 230$ MeV and polynomial
PL potential. Temperature values are given in MeV.} \label{fig3}
\end{figure}

\begin{figure}[hbt]
\includegraphics[width=1.\textwidth]{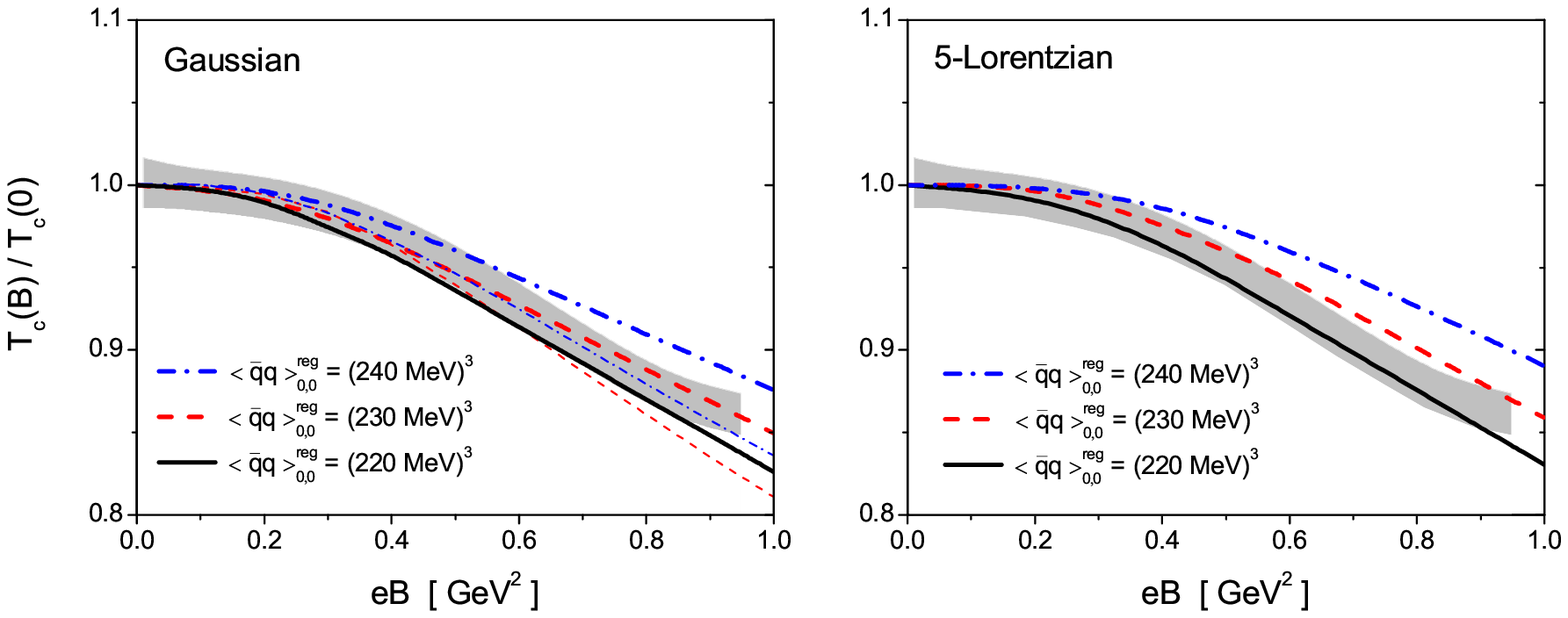}
\caption{Normalized critical temperatures as functions of $eB$ for various
model parametrizations. For comparison, LQCD results of
Ref.~\cite{Bali:2012zg} are indicated by the gray band. Left and right
panels correspond to Gaussian and 5-Lorentzian form factors, respectively.}
\label{fig4}
\end{figure}

Let us discuss the effect of the magnetic field on the phase transition
features. From Fig.~\ref{fig2} it is seen that the splitting between the
chiral restoration and deconfinement critical temperatures remains very
small in the presence of the external field (in fact, a detailed analysis
shows that the splitting gets reduced for larger values of $eB$). In
addition, it is seen that the nonlocal NJL models show inverse magnetic
catalysis. Indeed, contrary to what happens e.g.~in the standard local NJL
model~\cite{Kharzeev:2012ph,Andersen:2014xxa,Miransky:2015ava}, in our
models the chiral restoration critical temperature becomes lower as the
external magnetic field is increased. This is related to the fact that the
condensates do not show in general a monotonic increase with $B$ for a fixed
value of the temperature. The situation is illustrated in Fig.~\ref{fig3},
where we show the behavior of the averaged difference $\Delta \bar
\Sigma_{B,T}$ as a function of $eB$, for $T=0$ and for values of the
temperature in the critical region. The curves correspond to models with
Gaussian (left) and Lorentzian (right) form factors, $(-\langle\bar q q
\rangle^{\rm reg}_{0,0})^{1/3} = 230$ MeV, polynomial PL potential. For
these parametrizations, the critical temperatures for $B = 0$ are slightly
below 180 MeV (see Table~\ref{tab2}). While for $T=0$ the value of $\Delta
\bar \Sigma_{B,0}$ shows a monotonic growth with the external magnetic
field, it is seen that when the temperatures get closer to the critical
values the curves have a maximum and then start to decrease for increasing
$B$. This is the typical behavior associated to IMC and observed from
lattice QCD results, see e.g.~Fig.~2 of Ref.~\cite{Bali:2012zg}.
Qualitatively similar results are found for the other parametrizations in
Table I. Finally, in Fig.~\ref{fig4} we plot our results for the chiral
restoration critical temperatures $T_c(B)$, normalized to the corresponding
values at vanishing external magnetic field. The figure includes the curves
for nonlocal NJL models with Gaussian (left) and 5-Lorentzian (right) form
factors and different parameter sets (see the caption). The gray bands in both
panels show the results obtained in LQCD, taken from
Ref.~\cite{Bali:2012zg}. For comparison, for the Gaussian form factor we
have plotted with thin lines the results for the ``improved polynomial''.
Thick lines for both Gaussian and 5-Lorentzian form factors correspond to
the polynomial PL potential in Eq.~(\ref{upoly}). Results for the
logarithmic PL potential have been omitted, since (as stated above) the
transitions are found to be too steep in comparison with LQCD results. From
the figure it is clearly seen that the inverse magnetic catalysis effect is
observed for all considered parametrizations. In addition, for a given form
factor, the effect is found to be stronger for parameter sets leading to a
lower absolute value of the chiral quark condensates. As a general
conclusion, it can be stated that the behavior of the critical temperatures
with the external magnetic field is compatible with LQCD results, for
phenomenologically adequate values of the chiral condensate.

To shed some light on the mechanism that leads to the IMC effect in our
model it is worth noticing that the nonlocal form factor turns out to be a
function of the external magnetic field. This can be clearly seen from
Eq.~(\ref{mpk}). In addition, it is important to take into account that in
nonlocal NJL-like models the form factors play the role of some finite-range
gluon-mediated effective interaction. Thus, the magnetic field dependence of
the form factor can be understood as originated by the backreaction of the
sea quarks on the gluon fields. It is interesting to consider the effective
mass for the particular case of a Gaussian form factor, given by
Eq.~(\ref{gauss}). It can be seen that in this case the components of the
momentum that are parallel and transverse to the magnetic field become
disentangled. While for the $3,4$ components the original exponential form
$\exp{(-\bar p^2/\Lambda^2)}$ is maintained, the $1,2$ (transverse) part
leads to a factor given by a ratio of polynomials in $|q_f B|/\Lambda^2$,
which goes to zero for large $B$. In this way, for any value of $k$, the
strength of the effective coupling decreases as $B$ increases. This is
analogous to what happens with the $B$-dependent coupling constants
considered e.g.\ in Refs.~\cite{Ferreira:2014kpa,Farias:2014eca}, and thus
the IMC effect can be understood on these grounds.

\section{Summary \& conclusions}

We have studied the behavior of strongly interacting matter under a uniform
static external magnetic field in the context of a nonlocal chiral quark
model. In this approach, which can be viewed as an extension of the
Polyakov$-$Nambu$-$Jona-Lasinio model, the effective couplings between
quark-antiquark currents include nonlocal form factors that regularize
ultraviolet divergences in quark loop integrals and lead to a
momentum-dependent effective mass in quark propagators. We have worked out
the formalism introducing Ritus transforms of Dirac fields, which allow us to
obtain closed analytical expressions for the gap equations, the chiral quark
condensate, and the quark propagator. In addition, we have shown that these
expressions can also be obtained in the framework of a Schwinger-Dyson
approach.

We have considered the case of Gaussian and Lorentzian form factors,
choosing some sets of model parameters that allow us to reproduce the empirical
values of the pion mass and decay constants. At zero temperature, with these
parameterizations we have calculated the behavior of the subtracted flavor
average condensate $\Delta\bar\Sigma_{B,0}$ and the normalized condensate
difference $\Sigma^u_{B,0}-\Sigma^d_{B,0}$ as functions of the external
magnetic field $B$. Our results show the expected effect of magnetic
catalysis (condensates behave as growing functions of $B$), the curves
being in quantitative agreement with lattice QCD calculations with slight
dependence on the parametrization.

Finally we have extended the calculations to finite temperature systems,
including the couplings of fermions to the Polyakov loop. We have defined
chiral and PL susceptibilities in order to study the chiral restoration and
deconfinement transitions, which turn out to proceed as smooth crossovers
for the polynomial PL potential considered. From our numerical calculations,
on one hand it is seen that, for all considered values of $B$, both
transitions take place at approximately the same temperature, in agreement
with LQCD predictions. On the other hand, it is found that for temperatures
close to the transition region the subtracted flavor average condensate
$\Delta\bar\Sigma_{B,T}$ becomes a nonmonotonic function of $B$, which
eventually leads to the phenomenon of inverse magnetic catalysis, i.e., a
decrease of the critical temperature when the magnetic field gets increased.
This feature is also in qualitative agreement with LQCD expectations.
Moreover, for some parameterizations we find a remarkably good quantitative
agreement with the results from LQCD calculations for the behavior of the
normalized critical temperatures with $B$ (see Fig.~\ref{fig4}). The values
of the critical temperature at $T=0$, which show some dependence on the
parameterization and the PL potential, lie also within the range estimated
by LQCD results.

It is interesting to compare the nonlocal models with approaches in which
IMC is obtained by considering some {\em ad hoc} dependence of the effective
couplings on $B$ and/or $T$. The naturalness of the IMC behavior in our
framework can be understood by noticing that for a given Landau level the
associated nonlocal form factor turns out to be a function of the external
magnetic field, according to the convolution in Eq.~(\ref{mpk}). Since the
form factors can be identified with some gluon-mediated effective
interaction, the dependence on the magnetic field can be seen as originated
by the backreaction of the quarks on the gluon fields.

\section*{Acknowledgements}

This work has been supported in part by CONICET and ANPCyT (Argentina),
under grants PIP14-492, PIP12-449, and PICT14-03-0492, by the National
University of La Plata (Argentina), Project No.\ X718, by the Mineco
(Spain), under contract FPA2013-47443-C2-1-P, FPA2016-77177-C2-1-P, by the
Centro de Excelencia Severo Ochoa Programme, grant SEV-2014-0398, and by
Generalitat Valenciana (Spain), grant PrometeoII/2014/066.

\section*{Appendix A: Ritus eigenfunctions and Ritus transforms}

\newcounter{erasmo}
\renewcommand{\thesection}{\Alph{erasmo}}
\renewcommand{\theequation}{\Alph{erasmo}\arabic{equation}}
\setcounter{erasmo}{1} \setcounter{equation}{0} 

In this Appendix we provide the explicit form of the Ritus
eigenfunctions~\cite{Ritus:1978cj} and discuss some of the their properties.
These functions satisfy the eigenvalue equation
\begin{equation}
\Pi^2 \ \mathbb{E}_{\bar p} (x) \ = \ \epsilon_{\bar p} \ \mathbb{E}_{\bar
p} (x)\ ,
\label{ecautov}
\end{equation}
where, in accordance with the definition in the main text, $\Pi = -i
\rlap/\partial - q B x_1 \gamma_2$. Here, $\bar p=(k,p_2,p_3,p_4)$
represents the set of quantum numbers needed to label the eigenstates, the
eigenvalues of which are given by $\epsilon_{\bar p}= -(2k |q B| + p_3^2+p_4^2)$.
Working in Euclidean space and choosing the Weyl representation for the
Dirac matrices,
\begin{equation}
\vec \gamma=\begin{pmatrix} 0 & \vec \sigma
\\ - \vec \sigma & 0 \end{pmatrix}\ , \qquad \gamma_4= i \gamma_0 = i
\begin{pmatrix} 0 & \mathcal{I}  \\ \mathcal{I} & 0 \end{pmatrix} \ ,
\end{equation}
one has
\begin{equation}
\mathbb{E}_{\bar p} (x)\ = \ \sum_{\lambda=\pm} E_{\bar p \lambda}(x)\,
\Delta^\lambda\ ,
\label{ep}
\end{equation}
where $\Delta^{+}=\mbox{diag}(1,0,1,0)$, $\Delta^{-}=\mbox{diag}(0,1,0,1)$,
and
\begin{equation}
E_{\bar p\lambda}(x) \ = \ N_{k_\lambda} \
e^{i(p_2x_2+p_3x_3+p_4x_4)}\, D_{k_\lambda}(\rho)\ ,
\label{autofuncion}
\end{equation}
where $\rho = s \sqrt{2/|q B|} \, (q B \, x_1 - p_2)$, with $s=\mbox{sign}(q
B)$. The integer index $k_\lambda$ is related to the quantum number $k$ by
\begin{equation}
k_{\pm}= k - \frac{1}{2} \pm \frac{s}{2} \ ,
\label{a5}
\end{equation}
while $N_n = (4\pi|q B|)^{1/4}/\sqrt{n!}\,$. In
Eq.~(\ref{autofuncion}) we have introduced the cylindrical parabolic
functions defined by
\begin{equation}
D_n(x) \ = \ 2^{-n/2}\, e^{-x^2/4}\, H_n(x/\sqrt{2})\ ,
\label{dn}
\end{equation}
where $H_n(x)$ are the Hermite polynomials, with the standard convention
$H_{-1}(x)=0$. In fact, strictly speaking, for $k=0$ the Ritus eigenfunction
$\mathbb{E}_{\bar p}(x)$ should be defined as a $2\times 2$ matrix
\begin{equation}
\mathbb{E}_{(0,p_2,p_3,p_4)}(x) \ = \ (4\pi|q B|)^{1/4} \
e^{i(p_2x_2+p_3x_3+p_4x_4)}\, e^{-\rho^2/4}\;
\leavevmode\hbox{\small1\kern-3.8pt\normalsize1}_{(2\times 2)}\ ,
\end{equation}
where $\leavevmode\hbox{\small1\kern-3.8pt\normalsize1}_{(2\times 2)}$ is
the identity matrix in the subspace where $E_{\bar p \lambda}(x)$ is
nonzero. On the other hand, it is easily seen that the matrices
$\Delta^\lambda$ satisfy
\begin{eqnarray}
\Delta^\pm \Delta^{\pm} = \Delta^\pm\ ,
\quad
\Delta^\pm \Delta^{\mp} = 0\ ,
\quad
\Delta^\pm \gamma_{\perp} = \gamma_{\perp}\, \Delta^{\mp}\ ,
\quad
\Delta^\pm \gamma_{\parallel} = \gamma_{\parallel}\, \Delta^\pm\ ,
\label{propdel}
\end{eqnarray}
where $\gamma_\perp = (\gamma_1,\gamma_2)$ and $\gamma_\parallel =
(\gamma_3,\gamma_4)$.

As expected, along the direction of the magnetic field the function
$\mathbb{E}_{\bar p}(x)$ preserves the form of the energy eigenfunction of a
free particle, being labeled by a continuous index $p_3$ that corresponds
to the momentum component parallel to $\vec B$. This is also the situation
in the direction of the imaginary time. On the other hand, the quantum
numbers corresponding to the plane $x_1\,x_2$ depend on the gauge used to
describe the vector potential $A_\mu$. We have chosen the Landau gauge, for
which the states associated with the $x_1$ direction are quantized and
labeled by the integer index $k$. Along the $x_2$ direction, the
eigenfunction has the form of that of a free particle, with the
particularity that the eigenvalues do not depend on $p_2$, and hence the states
are degenerated. This last property leads to the useful relation
\begin{equation}
\int \frac{dp_2}{2\pi}  \; \mathbb{E}_{\bar p}(x)\,\mathbb{\bar E}_{\bar p}(x) \ =
\ \int \frac{dp_2}{2\pi}  \; \mathbb{\bar E}_{\bar p}(x)\,\mathbb{ E}_{\bar p}(x)
\ = \ |q B|\, P_{k,s}\ ,
\label{int2}
\end{equation}
where we have defined $\mathbb{\bar E}_{\bar p} = \gamma_0 \,
\mathbb{E}_{\bar p}^\dagger \, \gamma_0$ and $P_{k,\pm
1}=(1-\delta_{k0})\,\mathcal{I}+\delta_{k0}\,\Delta^\pm$. The operators
$P_{k,\pm 1}$ are projectors; i.e., they satisfy $P_{k,s} = (P_{k,s})^2$. It
is also seen that $P_{k,s} \; \mathbb{E}_{\bar p} =  \mathbb{E}_{\bar p} \;
P_{k,s} =\mathbb{E}_{\bar p}\,$.

The Ritus functions $\mathbb{E}_{\bar p}(x)$ satisfy orthonormality
and completeness relations, namely
\begin{eqnarray}
& & \int d^4x \; \mathbb{\bar E}_{\bar p}(x)\,\mathbb{ E}_{\bar p\,'}(x) \ =
\ \hat \delta_{\bar p,\bar p\,'}\, P_{k,s}\ ,
\\
& & \int \!\!\!\!\!\!\! \sum_{\bar p} \;
\mathbb{E}_{\bar p}(x)\,\mathbb{\bar E}_{\bar p}(x') \ = \ \delta^{(4)}(x-x')\ ,
\label{ritus}
\end{eqnarray}
where the following shorthand notations have been introduced
\begin{equation}
\int \!\!\!\!\!\!\! \sum _{\bar p} \equiv \frac{1}{2\pi} \sum_{k=0}^\infty \int
\frac{dp_2}{2\pi}\,\frac{dp_3}{2\pi}\,\frac{dp_4}{2\pi}\,\ ,
\quad
\hat \delta_{\bar p,\bar p\,'} \equiv (2\pi)^4\,\delta_{kk'}\,\delta(p_2-p'_2)
\,\delta(p_3-p'_3)\,\delta(p_4-p'_4)\ .
\label{defs}
\end{equation}
In addition, they satisfy the important identity
\begin{equation}
\Pi \; \mathbb{E}_{\bar p} (x)\ = \ \mathbb{E}_{\bar p} (x)\, \Big(\! -\! s\, \sqrt{2 k
|q B|}\ \gamma_2 + p_\parallel \cdot \gamma_\parallel \Big) \ ,
\label{impid}
\end{equation}
where $p_\parallel = (p_3,p_4)$.

Given the Ritus functions, one can define the Ritus transform of some
arbitrary Dirac function $\psi(x)$. One has
\begin{equation}
\psi_{\bar p} = \int d^4x  \ \mathbb{\bar E}_{\bar p}(x) \ \psi(x) \ ,
\qquad \bar \psi_{\bar p} = \int d^4x \ \bar \psi(x) \ \mathbb{E}_{\bar p}(x)\
,
\end{equation}
together with the inverse transforms
\begin{equation}
\psi(x) = \int  \!\!\!\!\!\!\! \sum _{\bar p}\; \mathbb{E}_{\bar p}(x) \ \psi_{\bar p}\ ,
\qquad \bar \psi(x) = \int  \!\!\!\!\!\!\! \sum _{\bar p}\; \bar \psi_{\bar p} \
\mathbb{\bar E}_{\bar p}(x)\ .
\end{equation}
In the same way, the Ritus transform $\mathcal{O}_{\bar p,\bar p\,'}$ of an
arbitrary operator $\mathcal{O}_{x,x'}$ satisfies
\begin{eqnarray}
\mathcal{O}_{\bar p,\bar p\,'} &=& \int d^4x \ d^4x' \ \
                       \mathbb{\bar E}_{\bar p}(x) \ \mathcal{O}_{x,x'} \ \mathbb{E}_{\bar p\,'}(x') \ , \label{Opp}
\\
\mathcal{O}_{x,x'} &=& \int \!\!\!\!\!\!\! \sum_{\;\,\bar p,\bar p\,'} \ \mathbb{E}_{\bar p}(x) \ \mathcal{O}_{\bar p,\bar p\,'}
\ \mathbb{\bar E}_{\bar p\,'}(x')\ .
\label{Oxx}
\end{eqnarray}

\section*{Appendix B: Details of the evaluation of $G^{\lambda,f}_{\bar p,\bar p\,'}$}

\newcounter{erasmoB}
\renewcommand{\thesection}{\Alph{erasmoB}}
\renewcommand{\theequation}{\Alph{erasmoB}\arabic{equation}}
\setcounter{erasmoB}{2} \setcounter{equation}{0} 

We start from the relation in Eq.~(\ref{Gpp}),
\begin{equation}
G^{\lambda,f}_{\bar p,\bar p\,'} =  \int d^4x \ d^4x' \ E^*_{\bar p\lambda} (x) \
\mathcal{G}(x-x') \, \exp\left[i \Phi_f(x,x')\right] \,
E_{\bar p\,'\lambda}(x')\ ,
\label{GppApp}
\end{equation}
where $\Phi_f(x,x')=  (q_f B/2) \, (x_2 - x_2')\, (x_1 +x_1')$, and the
functions $E_{\bar p\lambda}(x)$ are given in Eq.~(\ref{autofuncion}). To
work out this expression, we introduce the Fourier transform of
$\mathcal{G}(x)$,
\begin{eqnarray}
g(t^2) = \int d^4 x\ e^{-i t\, \cdot x} \; \mathcal{G}(x)\ ,
\end{eqnarray}
and perform the change of variables $x = z + y/2$, $x' = z - y/2$. In this
way, we get
\begin{equation}
G^{\lambda,f}_{\bar p,\bar p\,'} =  \int \frac{d^4 t}{(2\pi)^4}
\ g(t^2) \int d^4y \ d^4z \ E^*_{p\lambda} (z+y/2) \
\exp (i t \cdot y)  \, \exp(i q_f B y_2 z_1) \;
E_{p'\lambda}(z-y/2)\ .
\end{equation}
Given the explicit form of the functions $E_{\bar p\lambda}(x)$, the
integrals over $z_2,z_3,z_4$ and $y_3,y_4$ can be easily performed. We
obtain
\begin{eqnarray}
G^{\lambda,f}_{\bar p,\bar p\,'} &=&  (2 \pi)^3\, \delta(p_2-p_2')\,
\delta(p_3-p'_3)\, \delta(p_4-p'_4) \;
\Gamma^{\lambda,f}_{k,k',p_\parallel}\ ,
\label{gppp}
\end{eqnarray}
where
\begin{eqnarray}
\Gamma^{\lambda,f}_{k,k',p_\parallel} & = &
\ N_{k_\lambda} N_{k^\prime_\lambda}
\int \frac{d^2 t_\perp}{(2\pi)^2}
\ g(t_\perp^2 + p_\parallel^2) \int dz_1 d^2 y_\perp \nonumber \\
& &
\exp(-i p_2 y_2)\,
\exp(i t_\perp \cdot  y_\perp)\,
\exp(i q_f B y_2 z_1)\;
D_{k_\lambda}(\rho)\; D_{k'_\lambda}(\rho')\ ,
\end{eqnarray}
with $t_\perp = (t_1,t_2)$ and
\begin{eqnarray}
\rho = s_f \sqrt{\frac{2}{|q_f B|}} \; \big[q_f B \ (z_1+y_1/2) - p_2\big]\ ,
\quad
\rho' = s_f \sqrt{\frac{2}{|q_f B|}} \; \big[q_f B \ (z_1-y_1/2) - p_2\big]\ .
\end{eqnarray}
We recall here that $s_f=\mbox{sign}(q_f B)$, while $k_\lambda$ is related
to $k$ by Eq.~(\ref{a5}). We note now that the integration over $y_2$
introduces a factor $2\pi \, \delta(q_f B z_1- p_2 + t_2)$, which allows us to
easily perform the integral over $t_2$. Taking into account the explicit
form of $\rho$ and $\rho'$, we get
\begin{eqnarray}
\Gamma^{\lambda,f}_{k,k',p_\parallel}&=&
\frac{1}{\left[ 2\pi \; 2^{{k_\lambda}+{k'_\lambda}}\, {k_\lambda}!\, {k'_\lambda}! \right]^{1/2}}
\int d \gamma \ d \eta \ d \psi \
\ g\Big[\frac{|q_f B|}{2} (\gamma^2 + \eta^2)\, +\, p_\parallel^2\Big]\ \times
\nonumber \\
& & \exp(i \gamma \psi)\, \exp\Big(\!-\frac{\eta^2+\psi^2}{2}\Big)\;
H_{k_\lambda}\Big(\frac{\eta + \psi}{\sqrt{2}}\Big)\;
H_{k'_\lambda}\Big(\frac{\eta  - \psi}{\sqrt{2}}\Big)\ ,
\end{eqnarray}
where we have used the expression of $D_n$ in terms of Hermite polynomials,
Eq.~(\ref{dn}), and for convenience we have introduced the dimensionless
variables
\begin{equation}
\gamma = \sqrt{\frac{2}{|q_f B|}}\ t_1\ ,\qquad  \eta = s_f \sqrt{\frac{2}{|q_f B|}} \left( q_f B z_1 +
p_2\right)\ ,
\qquad \psi = \sqrt{\frac{|q_f B|}{2}} \ y_1\ .
\end{equation}
Making a new change of variables to polar coordinates $r,\phi$ in the
$\gamma\eta$ plane, we get
\begin{eqnarray}
\Gamma^{\lambda,f}_{k,k',p_\parallel}&=&
\int_0^\infty d r  \ r \
\ g\Big(\frac{|q_f B|}{2}\, r^2\, +\,p_\parallel^2\Big)\,
\exp(-r^2/2) \; I^\lambda_{k,k'}(r)\ ,
\label{gkk}
\end{eqnarray}
where
\begin{eqnarray}
I^\lambda_{k,k'}(r)&=&
\frac{1}{\left[ 2\pi \; 2^{{k_\lambda}+{k'_\lambda}}\, {k_\lambda}!\, {k'_\lambda}! \right]^{1/2}}
\int_0^{2\pi} d \phi \int_{-\infty}^\infty d \psi \
\exp\Big[-(\psi- i r \cos\phi)^2/2\,\Big]\ \times
\nonumber \\
& & \qquad \qquad\qquad\qquad\qquad H_{k_\lambda} \Big(\frac{r \sin\phi +
\psi }{\sqrt{2}}\Big)\; H_{k'_\lambda}\Big(\frac{r \sin\phi  -
\psi}{\sqrt{2}}\Big)\ .
\label{ikkp}
\end{eqnarray}
Next we carry out a translation into the complex plane $\psi$, namely $\psi
\rightarrow \psi' = \psi - i r \cos\phi$. Since the integrand in
Eq.~(\ref{ikkp}) is an analytic function, making use of Cauchy's theorem
one can show that the integration path can be taken along the ${\rm
Im}\,\psi\,' = 0$ axis. Thus, we obtain
\begin{eqnarray}
I^\lambda_{k,k'}(r)&=&
\frac{1}{\left[ 2\pi\; 2^{{k_\lambda}+{k'_\lambda}}\, {k_\lambda}!\,
{k'_\lambda}! \right]^{1/2}} \int_0^{2\pi} d \phi \int_{-\infty}^\infty d \psi
\;\exp(-\psi^2/2)
\nonumber \\
& & \qquad \qquad\qquad \qquad\quad H_{k_\lambda} \Big[\frac{ir \exp{(-i \phi)} +
\psi }{\sqrt 2}\Big]\; H_{k'_\lambda}\Big[\frac{-ir \exp{(i \phi)}  -
\psi}{\sqrt 2}\Big]\ .
\label{ikk}
\end{eqnarray}
Next, we use the relation $H_n(-x) = (-1)^n H_n (x)$ and the identity
(see Eq.~(7.377) of Ref.~\cite{grad})
\begin{eqnarray}
\int_{-\infty}^\infty dx\; e^{-x^2} H_m(x+y) \ H_n(x+z) \ = \ 2^n \;
\sqrt{\pi} \; m! \; z^{n-m} \; L^{n-m}_m(-2 y z)\ , \quad n\geq m\ ,
\label{hhlag}
\end{eqnarray}
where $L^a_b(x)$ are generalized Laguerre polynomials.
Finally, using
\begin{eqnarray}
\int_0^{2\pi} d \phi \ \exp(i \phi\, m) \ = \ 2 \pi\, \delta_{m0}\ ,
\end{eqnarray}
we obtain
\begin{equation}
I^\lambda_{k,k'}(r) \ = \ 2\pi\, (-1)^{k_\lambda}\, L_{k_\lambda}(r^2) \;
\delta_{kk'}\ .
\label{b14}
\end{equation}
Replacing Eq.~(\ref{b14}) in Eq.~(\ref{gkk}), and taking into account
Eq.~(\ref{gppp}), after a new change of variables $r \to |p_\perp| =
r\sqrt{|q_fB|/2}$ we end up with
\begin{equation}
G^{\lambda,f}_{\bar p,\bar p'} \ = \ \hat \delta_{\bar p,\bar p\,'} \
g^{\lambda,f}_{k,p_\parallel}\ ,
\end{equation}
where $g^{\lambda,f}_{k,p_\parallel}$ is given in Eq.~(\ref{mpk}).

\section*{Appendix C: Mean field quark propagator}

\newcounter{eraC}
\renewcommand{\thesection}{\Alph{eraC}}
\renewcommand{\theequation}{\Alph{eraC}\arabic{equation}}
\setcounter{eraC}{3} \setcounter{equation}{0} 

In this Appendix, we outline the derivation of the $u$ and $d$ quark
propagators within the MFA. We start from the two-point function in Ritus
space $\mathcal{D}^{\mbox{\tiny MFA},f}_{\bar p,\bar p\,'}$ which, as discussed in the main text,
is diagonal in Landau/momentum indices $\bar p$. The mean field
quark propagators in this space, for quark flavors $f=u,d$, are then given by
\begin{equation}
S^{\mbox{\tiny MFA},f}_{\bar p,\bar p\,'} \ = \ \big(\mathcal{D}^{\mbox{\tiny
MFA},f}_{\bar p,\bar p\,'}\big)^{-1} \ = \ \hat
\delta_{\bar p,\bar p\,'}\,\big( {\mathcal D}^{f}_{k,p_\parallel}
\big)^{-1}\ ,
\label{}
\end{equation}
with ${\mathcal D}^{f}_{k,p_\parallel}$ given by Eq.~(\ref{twopoint}). Since
this operator is nondiagonal only in Dirac space, it can be easily inverted.
Defining ${\cal S}^{f}_{k,p_\parallel} = ( {\mathcal D}^{f}_{k,p_\parallel}
)^{-1}$, one finds that ${\cal S}^{f}_{k,p_\parallel}$ can be written as
\begin{equation}
{\cal S}^{f}_{k,p_\parallel} \ = \ \sum_{\lambda=\pm} \Big[\hat
A^{\lambda,f}_{k,p_\parallel} - \hat B^{\lambda,f}_{k,p_\parallel}\,
p_\parallel\cdot\gamma_\parallel + s_f\sqrt{2kB_f}
\big(\hat C^{\lambda,f}_{k,p_\parallel} - \hat D^{\lambda,f}_{k,p_\parallel}\,
p_\parallel\cdot\gamma_\parallel\big)\gamma_2\Big]\Delta^\lambda \ ,
\end{equation}
where we have defined $B_f = |q_fB|$, and the functions $\hat
A^{\lambda,f}_{k,p_\parallel}$ to $\hat D^{\lambda,f}_{k,p_\parallel}$ are
given in Eqs.~(\ref{bb}-\ref{aa}). Notice that in the particular case $k=0$
(i.e.\ $k_\lambda = 0$ or $-1$) the Dirac space is reduced to a
two-dimensional one; therefore, only the coefficients $\hat
A^{\lambda,f}_{k,p_\parallel}$ and $\hat B^{\lambda,f}_{k,p_\parallel}$ with
$k_\lambda = 0$ need to be considered.

To find the expression for the propagator in coordinate space, we have to
perform the Ritus antitransform of $S^{\mbox{\tiny MFA},f}_{\bar p,\bar p\,'}$. One has
\begin{eqnarray}
\!\!\!\!\!\!\!\!\!\!\!\!\!
S^{\mbox{\tiny MFA},f}_{x,x'} & = & \int \!\!\!\!\!\!\! \sum_{\;\;\bar p,\bar p\,'}\;\mathbb{E}_{\bar p} (x)\,
S^{\mbox{\tiny MFA},f}_{\bar p,\bar p\,'}\; \bar{\mathbb{E}}_{\bar p\,'} (x')
\nonumber \\
\!\!\!\!\!\!\!\! & = & \frac{1}{2\pi}\sum_{k=0}^\infty \int
\frac{d^2p_\parallel}{(2\pi)^2}\, e^{i p_\parallel\cdot \Delta x_\parallel}\sum_{\lambda,\lambda'=\pm}
I^{\lambda\lambda'}
\,\Big[ \delta_{\lambda\lambda'}
 \big(\hat A^{\lambda,f}_{k,p_\parallel} - \hat B^{\lambda,f}_{k,p_\parallel}\,
p_\parallel\cdot\gamma_\parallel\big)
\Delta^\lambda\,  \nonumber\\
& & \qquad \qquad + s_f \sqrt{2kB_f}\, (1-\delta_{\lambda\lambda'})  \big(
\hat C^{\lambda',f}_{k,p_\parallel} - \hat D^{\lambda',f}_{k,p_\parallel}\,
p_\parallel\cdot\gamma_\parallel\big)\gamma_2\,\Delta^{\lambda'} \Big]\ ,
\label{sxx}
\end{eqnarray}
where we have defined $\Delta x_\parallel =  (\Delta x_3,\Delta x_4)$, with $\Delta x_i = x_i - x'_i$, and the integrals
$I^{\lambda\lambda'}$ are given by
\begin{equation}
I^{\lambda\lambda'} \ = \ N_{k_\lambda}\,N_{k_{\lambda'}} \int \frac{dp_2}{2\pi}\;
e^{ip_2(x_2-x_2')}\,D_{k_\lambda}(\rho)\,D_{k_{\lambda'}}(\rho')\ ,
\label{idef}
\end{equation}
with $\rho^{(\prime)} = s_f\sqrt{2/B_f}[\,q_f B\, x_1^{(\prime)}-p_2] =
\sqrt{2B_f}[\,x_1^{(\prime)}-(s_f/B_f)\,p_2]$. Let us analyze separately the
integrals $I^{\pm\pm}$ and $I^{\pm\mp}$. Considering the explicit
expressions for $N_{k_\lambda}$ and $D_{k_\lambda}(x)$ [see Eq.~(\ref{dn})],
and performing the translation $p_2 = q_2 + s_f B_f(x_1+x_1')/2$, one has
\begin{eqnarray}
I^{\lambda\lambda} & = & \sqrt{\frac{B_f}{\pi}}\,
\frac{2^{-k_\lambda}}{k!}\; \exp[i\Phi_f(x,x')]\,
\exp(-B_f \Delta x_1^2/4) \int_{-\infty}^\infty dq_2\;\exp(iq_2 \Delta x_2)\,
\times
\nonumber \\
& & \exp(-q_2^2/B_f)\;H_{k_\lambda}\Big(\frac{\sqrt{B_f}\Delta x_1}{2} - \frac{s_f
q_2}{\sqrt{B_f}}\Big)
\; H_{k_\lambda}\Big(-\frac{\sqrt{B_f}\Delta x_1}{2} - \frac{s_f
q_2}{\sqrt{B_f}}\Big)\ ,
\end{eqnarray}
where $\Phi_f(x,x')$ is the already defined Schwinger phase. Now it is
possible to carry out a translation in the complex plane to a new variable
$\omega = (q_2 -i B_f\Delta x_2/2)s_f/\sqrt{B_f}$. Since the integrand is an
analytic function in the whole plane, the integral can be calculated along
the ${\rm Im}\,\omega = 0$ axis. One gets in this way
\begin{eqnarray}
I^{\lambda\lambda} & = & \frac{B_f}{\sqrt{\pi}}\,
\frac{2^{-k_\lambda}}{k!}\; \exp[i\Phi_f(x,x')]\, \exp(-B_f \Delta
x_\perp^2/4) \int_{-\infty}^\infty d\omega\;\exp(-\omega^2) \,\times
\nonumber \\
& & H_{k_\lambda}\Big[\omega - \sqrt{B_f}(\Delta x_1-i s_f\Delta
x_2)/2\Big] \; H_{k_\lambda}\Big[\omega + \sqrt{B_f}(\Delta x_1 +
i s_f\Delta x_2)/2\Big] \ ,
\label{ill}
\end{eqnarray}
where $\Delta x_\perp = (\Delta x_1, \Delta x_2)$. The integral in
Eq.~(\ref{ill}) can be evaluated using the relation in Eq.~(\ref{hhlag}),
which leads to
\begin{equation}
I^{\lambda\lambda} \ = \ B_f\,
\exp[i\Phi_f(x,x')]\, \exp[-B_f \Delta x_\perp^2/4]
\; L_{k_\lambda}\big(B_f \Delta x_\perp^2/2\big) \ .
\label{illcalc}
\end{equation}
Next, let us consider the integral
\begin{equation}
K^{(0)}(m,y_\perp) \ = \ \int d^2p_\perp \; e^{i p_\perp\cdot y_\perp}\;
\exp(-p_\perp^2/ B_f)\; L_m(2p_\perp^2/B_f)\ ,
\label{i1}
\end{equation}
where $p_\perp = (p_1,p_2)$, $y_\perp = (y_1,y_2)$. One has
\begin{eqnarray}
K^{(0)}(m,y_\perp) & = & \int_0^\infty d\left|p_\perp\right| \; \left|p_\perp\right|\, \exp(-p_\perp^2/ B_f)\,
L_m(2p_\perp^2/B_f)\;\int_0^{2\pi} d\theta \; e^{i\left|p_\perp\right|(y_1\cos\theta +
y_2\sin \theta)} \nonumber \\
& = & 2\pi \int_0^\infty d\left|p_\perp\right| \; \left|p_\perp\right|\, \exp(-p_\perp^2/ B_f)\,
L_m(2p_\perp^2/B_f)\; J_0(\left|p_\perp\right| \left|y_\perp\right|) \nonumber \\
& = & \pi B_f\, (-1)^m \, \exp(-B_f\, y_\perp^2/4)\, L_m(B_f\, y_\perp^2/2) \ ,
\label{i1calc}
\end{eqnarray}
where $J_0(x)$ is a Bessel function. The last equality in Eq.~(\ref{i1calc})
has been obtained using the following general relation, which involves
generalized Laguerre polynomials and Bessel functions:
\begin{equation}
\int_0^\infty dx\, x^{\nu+1}\ e^{-\beta x^2} L_m^\nu(\alpha x^2) J_\nu(x y)
= (2\beta)^{-\nu-1}\Big(1-\frac{\alpha}{\beta}\Big)^m y^\nu e^{-y^2/(4\beta)}\,
L_m^\nu\bigg[\frac{\alpha y^2}{4\beta(\alpha-\beta)}\bigg]\ .
\label{gen}
\end{equation}
{}From Eqs.~(\ref{illcalc}), (\ref{i1}), and (\ref{i1calc}), we end up with
\begin{eqnarray}
I^{\lambda\lambda} & = & \frac{1}{\pi}\,
\exp[i\Phi_f(x,x')]\,(-1)^{k_\lambda}\,
K^{(0)}(k_\lambda,\Delta x_\perp)\nonumber \\
& = & 4\pi\, \exp[i\Phi_f(x,x')]\, (-1)^{k_\lambda}\! \int
\frac{d^2p_\perp}{(2\pi)^2}\; e^{i p_\perp \cdot\Delta x_\perp}
\; \exp(-p_\perp^2/ B_f)\; L_{k_\lambda}(2p_\perp^2/B_f)\ .
\label{illfin}
\end{eqnarray}

A similar procedure can be carried out for the calculation of the integrals
$I^{\pm\mp}$. Performing the same changes of variables as in the previous
case, we obtain
\begin{eqnarray}
\hspace{-0.5cm} I^{\pm\mp} & = & \frac{B_f}{\sqrt{\pi}}\,
\frac{2^{-(k_+ + k_-)/2}}{\sqrt{k}\,(k-1)!}\; \exp[i\Phi_f(x,x')]\, \exp(-B_f \Delta x_\perp^2/4)\,
(-1)^{k_+ + k_-} \int_{-\infty}^\infty d\omega\;e^{-\omega^2} \,\times
\nonumber \\
& & H_{k_+}\bigg[\omega \mp \frac{\sqrt{B_f}}{2}(\Delta x_1 \mp i s_f\Delta
x_2)\bigg] \; H_{k_-}\bigg[\omega \pm \frac{\sqrt{B_f}}{2}(\Delta x_1
\pm i s_f\Delta x_2)\bigg] \nonumber \\
& = & B_f\sqrt{\frac{B_f}{2k}}\, s_f \exp[i\Phi_f(x,x')] (\pm \Delta x_1 -i\Delta x_2)
\exp\!\Big(\!-\frac{B_f \Delta x_\perp^2}{4}\Big) L^1_{k-1}\Big(\frac{B_f
\Delta x_\perp^2}{2}\Big) \, ,
\label{imm}
\end{eqnarray}
where we have used once again the relation in Eq.~(\ref{hhlag}) to evaluate
the integral over $\omega$. Notice that for $k=0$ one has $I^{+-} =
I^{-+}=0$ automatically from the definition in Eq.~(\ref{idef}), since
either $k_+=-1$ or $k_-=-1$, and $D_{-1}(\rho^{(\prime)}) =0$. Now, let us consider
the integrals
\begin{equation}
K_j^{(1)}(m,y_\perp) \ = \ \int d^2p_\perp \; p_j\; e^{i p_\perp \cdot y_\perp}\;
\exp(-p_\perp^2/ B_f) \; L_m^1(2p_\perp^2/B_f)\ ,
\label{i2}
\end{equation}
where $j=1,2$. Using Eq.~(\ref{gen}) with $\nu = 1$, it is easy to show that
\begin{eqnarray}
K_j^{(1)}(m,y_\perp) & = & 2\pi\, i\, \frac{y_j}{\left|y_\perp\right|} \int_0^\infty
d\left|p_\perp\right| \; p_\perp^2 \, \exp(-p_\perp^2/ B_f)\,
L_m^1(2p_\perp^2/B_f)\; J_1( \left|p_\perp\right| \left|y_\perp\right|) \nonumber \\
& = & \frac{\pi}{2}\, i\, B_f^2\, (-1)^m \, y_j\, \exp(-B_f\, y_\perp^2/4)\,
L_m^1(B_f\, y_\perp^2/2) \ ,
\label{i2calc}
\end{eqnarray}
from which we get
\begin{eqnarray}
I^{\pm\mp} & = & (-i)\,\frac{2}{\pi}\,
s_f\,\exp[i\Phi_f(x,x')]\, \frac{(-1)^k}{\sqrt{2kB_f}} \Big[\!\mp K_1^{(1)}(k-1,\Delta x_\perp)
+ i K_2^{(1)}(k-1,\Delta x_\perp)\,\Big]\nonumber \\
& = & - i\, 8\pi\,
s_f\,\exp[i\Phi_f(x,x')]\, \frac{(-1)^k}{\sqrt{2kB_f}} \int
\frac{d^2p_\perp}{(2\pi)^2} \; e^{i \Delta x_\perp\cdot p_\perp}\ \times
\nonumber \\
& & \qquad\qquad \qquad\qquad \qquad\qquad\quad
(\mp p_1 + i p_2)\, \exp(-p_\perp^2/ B_f)\; L^1_{k-1}(2p_\perp^2/B_f)\ .
\label{immfin}
\end{eqnarray}

The results in Eqs.~(\ref{illfin}) and (\ref{immfin}) can be put together as
\begin{eqnarray}
I^{\lambda\lambda'} & = & 4\pi\,(-i)^{k_\lambda +
k_{\lambda'}}\bigg(\frac{2}{\sqrt{2k\,B_f}}\bigg)^{|k_\lambda -
k_{\lambda'}|}
\,\exp[i\Phi_f(x,x')]\int
\frac{d^2p_\perp}{(2\pi)^2} \; e^{i \Delta x_\perp\cdot p_\perp}\, \exp(-p_\perp^2/ B_f)
\ \times \nonumber \\
& & \Big[(k_\lambda - k_{\lambda'}\!)\,p_1- is_fp_2\Big]^{|k_\lambda - k_{\lambda'}|}
\, L_{(k_\lambda + k_{\lambda'}-|k_\lambda - k_{\lambda'}|)/2}^{|k_\lambda -
k_{\lambda'}|}(2p_\perp^2/B_f)\
\end{eqnarray}
(notice that an analogous expression has been obtained in
Ref.~\cite{Watson:2013ghq}). Replacing into Eq.~(\ref{sxx}), and noting that
$-i(\pm p_1+ip_2)\gamma_2\Delta^\pm = p_\perp\cdot \gamma_\perp
\Delta^\pm\,$, we finally arrive at
\begin{equation}
S^{\mbox{\tiny MFA},f}_{x,x'} \ = \ \exp[i\Phi_f(x,x')]\,\int \frac{d^4p}{(2\pi)^4}\
e^{i\, p\cdot (x-x')}\, \tilde S^f(p_\perp,p_\parallel)\ ,
\end{equation}
where
$\tilde S^f (p_\perp,p_\parallel)$ is given by Eq.~(\ref{sfp}).

\section*{Appendix D: Derivation of the gap equation using the Schwinger-Dyson formalism}

\newcounter{eraD}
\renewcommand{\thesection}{\Alph{eraD}}
\renewcommand{\theequation}{\Alph{eraD}\arabic{equation}}
\setcounter{eraD}{4} \setcounter{equation}{0} 

In this Appendix we derive the gap equation using the Schwinger-Dyson (SD)
formalism discussed, e.g., in
Refs.~\cite{Leung:1996qy,Watson:2013ghq,Mueller:2014tea}. We start by
considering an interaction term of the form
\begin{equation}
S^{\rm int}_E \ = \ -\frac{1}{2} \int d^4 x_1\, d^4 x_2\, d^4 x_3\, d^4 x_4 \;
K_{\gamma_1,\gamma_2,\gamma_3,\gamma_4}(x_1,x_2,x_3,x_4) \; \bar
\psi_{\gamma_1}(x_1)\, \psi_{\gamma_2}(x_2)\, \bar \psi_{\gamma_3}(x_3)\,
\psi_{\gamma_4}(x_4) \ ,
\end{equation}
where $\gamma_i$ stands for a set of Dirac and internal indexes (i.e.\ color
and flavor). The corresponding SD equation for the two-point function in the
Hartree approximation is
\begin{equation}
\big( D_{x,x'} \big)_{\alpha,\beta} = \big( D^{(0)}_{x,x'}
\big)_{\alpha,\beta}   + \int d^4 x_3\,  d^4 x_4 \;
K_{\alpha,\beta,\gamma_3,\gamma_4}(x,x',x_3,x_4) \left( S_{x_4,x_3}
\right)_{\gamma_4,\gamma_3}\ ,
\end{equation}
where $D^{(0)}_{x,x'}$ is the free two-point function and $S_{x,x'}$ is the effective quark propagator.

The explicit form of the interaction kernel
$K_{\gamma_1,\gamma_2,\gamma_3,\gamma_4}(x_1,x_2,x_3,x_4)$ for the case we
are interested in can be read off from Eq.~(\ref{cuOGE}). Taking into
account that, due to the nonlocal character of the interaction, the coupling
with a gauge field requires the replacement in Eq.~(\ref{transport}), for
our nonlocal model in the presence of an external field we have
\begin{eqnarray}
K_{\gamma_1,\gamma_2,\gamma_3,\gamma_4}(x_1,x_2,x_3,x_4) & = &
G\; {\cal G}(x_1-x_2) \; {\cal G}(x_3-x_4) \; \delta^{(4)}\left(\bar x_{12}-\bar x_{34}\right)
\, \times \nonumber \\
& & \hspace{-4.3cm} \big( \gamma_0 \mathcal{W}\left(  x_1, \bar x_{12} \right)
\gamma_0 \, \Gamma_a \, \mathcal{W}\left( \bar  x_{12}, x_{2} \right)
\big)_{\gamma_1,\gamma_2} \big( \gamma_0 \mathcal{W}\left(  x_3, \bar
x_{34} \right) \gamma_0\, \Gamma_a \, \mathcal{W}\left( \bar  x_{34}, x_{4}
\right) \big)_{\gamma_3,\gamma_4}\ ,
\end{eqnarray}
where $\bar x_{ij} = (x_i + x_j)/2$. Replacing this expression in the SD
equation above, and considering the particular case of a constant magnetic
field along the 3-axis, in the Landau gauge we have
\begin{eqnarray}
D^f_{x,x'} & = & D^{(0),f}_{x,x'}\, + \, G  \; {\cal G}(x-x') \, \exp[ i
\Phi_f(x,x')]\, \times\nonumber \\
& & N_c \, \int d^4 y\,  d^4 y' \ {\cal G}(y-y')  \; \delta^{(4)}(\bar x
-\bar y)\, \sum_{f'=u,d}\; \mbox{tr}_D \left\{ \exp\big[ i \Phi_f (y,y')\big] \;
S^{f'}_{y',y} \right\}\ ,
\label{sdeq1}
\end{eqnarray}
where $\bar x = (x+x')/2$, $\bar y = (y+y')/2$, and $\Phi_f(x,x')$ is the
Schwinger phase introduced in Eq.~(\ref{schwingerphase}). We have assumed that,
due to parity conservation, only $\Gamma_0 =
\leavevmode\hbox{\small1\kern-3.8pt\normalsize1}$ is relevant at this level.
Thus, the solution of the SD equation has to be diagonal in flavor space,
allowing us to write the two-point function (and the corresponding propagator)
as in Eq.~(\ref{diago}). Note that in Eq.~(\ref{sdeq1}) the symbol
$\mbox{tr}_D$ stands for the trace in Dirac space, since the traces in color
and flavor spaces have already been taken.

To proceed we use the well-known fact (see e.g. Ref.~\cite{Watson:2013ghq})
that the two-point function of a free fermion in an external magnetic field
is given (in Euclidean space) by
\begin{equation}
D^{(0),f}_{x,x'} \ = \
\exp[i \Phi^f(x,x')] \int \frac{d^4 p}{(2\pi)^4}  \ e^{i p \cdot (x-x')}
\left( \rlap/p + m_c \right)
\label{d5}
\end{equation}
Replacing this relation into Eq.~(\ref{sdeq1}), we see that the rhs of the
resulting equation can be written as the product of a Schwinger phase factor
times a translational invariant function (i.e.\ a function that depends only
on $x-x'$). Thus, this has to be the form of $D^f_{x,x'}\,$. A suitable ansatz
for the Dirac structure of a two-point function of this type has been given
in Ref.~\cite{Watson:2013ghq}. Using our notation and conventions, its Ritus
transform reads
\begin{equation}
D^{f}_{k,p_\parallel} \ = \ \sum_{\lambda=\pm} \left[ A^{\lambda,f}_{k,p_\parallel} +
B^{\lambda,f}_{k,p_\parallel} \, p_\parallel \cdot \gamma_\parallel - s_f
\sqrt{2 k B_f} \left( C^{\lambda,f}_{k,p_\parallel} +
D^{\lambda,f}_{k,p_\parallel} \, p_\parallel \cdot \gamma_\parallel \right)
\gamma_2 \right] \Delta^\lambda \ .
\label{dapp}
\end{equation}

The Ritus transform of the associated propagator can be obtained by
inverting this $4\times 4$ matrix. It can be expressed as
\begin{equation}
S^{f}_{k,p_\parallel} = \sum_{\lambda=\pm}
\left[ \hat A^{\lambda,f}_{k,p_\parallel} - \hat B^{\lambda,f}_{k,p_\parallel} \ p_\parallel \cdot \gamma_\parallel
+ s_f \sqrt{2 k B_f}
\left( \hat C^{\lambda,f}_{k,p_\parallel} - \hat D^{\lambda,f}_{k,p_\parallel} \ p_\parallel \cdot \gamma_\parallel
\right) \gamma_2 \right] \, \Delta^\lambda\ ,
\label{sf-sd}
\end{equation}
where
\begin{eqnarray}
\hat A^{\pm,f}_{k,p_\parallel} & = & \frac{A^{\mp,f}_{k,p_\parallel} \
\Delta_1 \pm p^2_\parallel \ B^{\mp,f}_{k,p_\parallel}  \
\Delta_2}{\Delta}\ , \nonumber \\
\hat B^{\pm,f}_{k,p_\parallel} & = & \frac{B^{\mp,f}_{k,p_\parallel} \
\Delta_1 \mp \ A^{\mp,f}_{k,p_\parallel}  \ \Delta_2}{\Delta}\ , \nonumber \\
\hat C^{\pm,f}_{k,p_\parallel} & = & \frac{C^{\mp,f}_{k,p_\parallel} \
\Delta_1 \pm p^2_\parallel \ D^{\pm,f}_{k,p_\parallel}  \ \Delta_2}{\Delta}\ ,
\nonumber \\
\hat D^{\pm,f}_{k,p_\parallel} & = & - \frac{D^{\mp,f}_{k,p_\parallel} \
\Delta_1 \mp \ C^{\mp,f}_{k,p_\parallel}  \ \Delta_2}{\Delta} \ ,
\label{hatAD}
\end{eqnarray}
with the definitions
\begin{eqnarray}
\Delta_1 &=& A^{+,f}_{k,p_\parallel}  A^{-,f}_{k,p_\parallel}  +
p_\parallel^2 \ B^{+,f}_{k,p_\parallel}  B^{-,f}_{k,p_\parallel}  + 2 k B_f
\left( C^{+,f}_{k,p_\parallel}  C^{-,f}_{k,p_\parallel}  + p_\parallel^2 \
D^{+,f}_{k,p_\parallel}  D^{-,f}_{k,p_\parallel}  \right)
\ ,\nonumber \\
\Delta_2 &=& A^{+,f}_{k,p_\parallel}  B^{-,f}_{k,p_\parallel}  -
B^{+,f}_{k,p_\parallel}  A^{-,f}_{k,p_\parallel}  + 2 k B_f \left(
C^{+,f}_{k,p_\parallel}  D^{+,f}_{k,p_\parallel}  - C^{-,f}_{k,p_\parallel}
D^{-,f}_{k,p_\parallel}  \right)\ , \nonumber \\
\Delta & = & \Delta_1^2 + p_\parallel^2 \ \Delta_2^2\ .
\end{eqnarray}
The particular value $k=0$ should be considered separately. In this case the
above relations for $A^{\lambda,f}_{k,p_\parallel}$ and
$B^{\lambda,f}_{k,p_\parallel}$ simplify to
\begin{equation}
\hat A^{\lambda,f}_{0,p_\parallel} \ = \ \frac{A^{\lambda,f}_{0,p_\parallel}}
{{A^{\lambda,f}_{0,p_\parallel}}^2 + p_\parallel^2\,{B^{\lambda,f}_{0,p_\parallel}}^2}
\ \ , \qquad
\hat B^{\lambda,f}_{0,p_\parallel} \ = \ \frac{B^{\lambda,f}_{0,p_\parallel}}
{{A^{\lambda,f}_{0,p_\parallel}}^2 + p_\parallel^2\,{B^{\lambda,f}_{0,p_\parallel}}^2}
\ \ ,
\end{equation}
while $\hat C^{\lambda,f}_{0,p_\parallel}$ and $\hat
D^{\lambda,f}_{0,p_\parallel}$ are multiplied by zero in Eq.~(\ref{sf-sd}),
and need not be defined.

Following the same steps as those sketched in App.~C it can be shown that
the two-point function and the quark propagator in coordinate space can be
written as
\begin{eqnarray}
D^f_{x,x'} &=& \exp[i \Phi_f(x,x')] \int \frac{d^4 p}{(2\pi)^4}  \ e^{i p
\cdot (x-x')} \ \tilde D^f(p_\perp,p_\parallel)\ ,
\nonumber \\
S^f_{x,x'} &=& \exp[i \Phi_f(x,x')] \int \frac{d^4 p}{(2\pi)^4}  \ e^{i p
\cdot (x-x')} \ \tilde S^f(p_\perp,p_\parallel)\ .
\label{d6}
\end{eqnarray}
The functions $\tilde D^f(p_\perp,p_\parallel)$ and $\tilde
S^f(p_\perp,p_\parallel)$ are given by
\begin{eqnarray}
\!\!\!\!\!\tilde D^f(p_\perp,p_\parallel) &=& \sum_{\lambda=\pm} \left[
a^{\lambda,f}_{p_\perp,p_\parallel} + b^{\lambda,f}_{p_\perp,p_\parallel} \
p_\parallel \cdot \gamma_\parallel + \left(
c^{\lambda,f}_{p_\perp,p_\parallel} + d^{\lambda,f}_{p_\perp,p_\parallel} \
p_\parallel \cdot \gamma_\parallel \right) p_\perp \cdot \gamma_\perp
\right] \Delta^\lambda\ ,
\nonumber \\
\!\!\!\!\!\tilde S^f(p_\perp,p_\parallel) &=& \sum_{\lambda=\pm} \left[ \hat
a^{\lambda,f}_{p_\perp,p_\parallel} - \hat
b^{\lambda,f}_{p_\perp,p_\parallel} \ p_\parallel \cdot \gamma_\parallel +
\left( - \hat c^{\lambda,f}_{p_\perp,p_\parallel} + \hat
d^{\lambda,f}_{p_\perp,p_\parallel} \ p_\parallel \cdot \gamma_\parallel
\right) p_\perp \cdot \gamma_\perp \right] \Delta^\lambda \ ,
\label{d7}
\end{eqnarray}
where the functions $a^{\lambda,f}_{p_\perp,p_\parallel}, \dots$ are related
to $A^{\lambda,f}_{k,p_\parallel},\dots$ through
\begin{eqnarray}
\left(
  \begin{array}{c}
     a^{\lambda,f}_{p_\perp,p_\parallel} \\
     b^{\lambda,f}_{p_\perp,p_\parallel} \\
  \end{array}
\right)
&=& 2 e^{-p_\perp^2/B_f} \sum_{k=0}^\infty (-1)^{k_\lambda} \ L_{k_\lambda}\left(2 p_\perp^2/B_f\right) \
\left(
  \begin{array}{c}
     A^{\lambda,f}_{k,p_\parallel} \\
     B^{\lambda,f}_{k,p_\parallel} \\
  \end{array}
\right)\ ,
\nonumber \\
& & \nonumber \\
\left(
  \begin{array}{c}
     c^{\lambda,f}_{p_\perp,p_\parallel} \\
     d^{\lambda,f}_{p_\perp,p_\parallel} \\
  \end{array}
\right)
&=& 4 e^{-p_\perp^2/B_f} \sum_{k=1}^\infty (-1)^{k-1} \ L^1_{k-1}\left(2 p_\perp^2/B_f\right) \
\left(
  \begin{array}{c}
     C^{\lambda,f}_{k,p_\parallel} \\
     D^{\lambda,f}_{k,p_\parallel} \\
  \end{array}
\right)\ ,
\label{d8}
\end{eqnarray}
and similar relations hold for the functions $\hat
a^{\lambda,f}_{p_\perp,p_\parallel}$, $\hat A^{\lambda,f}_{k,p_\parallel}$,
etc.\ in the expression of the propagator. Note that using the orthogonality
of generalized Laguerre polynomials (see, e.g., Eq.~(3) of Sec.~7.414 in
Ref.~\cite{grad}),
\begin{equation}
\int_0^\infty dx\; x^\alpha e^{-x} \; L_n^\alpha(x) \; L_m^\alpha(x) \ = \
\frac{\Gamma(\alpha + n +1)}{n!} \ \delta_{nm} \ , \qquad
\mbox{Re}(\alpha) > 0\ ,
\end{equation}
these relations can be inverted to give
\begin{eqnarray}
\left(
  \begin{array}{c}
     A^{\lambda,f}_{k,p_\parallel} \\
     B^{\lambda,f}_{k,p_\parallel} \\
  \end{array}
\right)
&=& \frac{4 \pi}{B_f} (-1)^{k_\lambda} \int \frac{ d^2 p_\perp}{(2 \pi)^2}\
e^{-p_\perp^2/B_f}  \; L_{k_\lambda}(2 p_\perp^2/B_f) \,
\left(
  \begin{array}{c}
     a^{\lambda,f}_{p_\perp,p_\parallel} \\
     b^{\lambda,f}_{p_\perp,p_\parallel} \\
  \end{array}
\right)\ ,
\nonumber \\
& & \nonumber \\
\left(
  \begin{array}{c}
     C^{\lambda,f}_{k,p_\parallel} \\
     D^{\lambda,f}_{k,p_\parallel} \\
  \end{array}
\right)
&=&\frac{4 \pi}{B_f^2}\; \frac{(-1)^{k-1}}{\!\!k} \int \frac{ d^2 p_\perp}{(2 \pi)^2}\
p_\perp^2\; e^{-p_\perp^2/B_f}\; L^1_{k-1}(2 p_\perp^2/B_f) \, \left(
  \begin{array}{c}
     c^{\lambda,f}_{p_\perp,p_\parallel} \\
     d^{\lambda,f}_{p_\perp,p_\parallel} \\
  \end{array}
\right)\ .
\label{d10}
\end{eqnarray}

We can now go back to the SD equation, Eq.~(\ref{sdeq1}). Using
Eqs.~(\ref{d5}) and (\ref{d6}) we have
\begin{equation}
\tilde D^f(p_\perp,p_\parallel) \ = \ \rlap/p + m_c + G \, N_c \, g(p^2)\; \sum_{f'=u,d}
\; \int \frac{d^4 q}{(2 \pi)^4} \ g(q^2)\; \mbox{tr}_D\! \left[ \tilde S^{f'}\!(q_\perp,q_\parallel)
\right]\ .
\end{equation}
Taking into account the explicit form of $\tilde D^f_{p}$ and $\tilde
S^f_{p}$ given by Eq.~(\ref{d7}), it is seen that the functions entering
$\tilde D^f(p_\perp,p_\parallel)$ should satisfy
\begin{eqnarray}
a^{\lambda,f}_{p_\perp,p_\parallel} = m_c + \bar \sigma \, g(p^2)\ ,
\qquad
b^{\lambda,f}_{p_\perp,p_\parallel} = c^{\lambda,f}_{p_\perp,p_\parallel} = 1
\ ,\qquad
d^{\lambda,f}_{p_\perp,p_\parallel} = 0\ ,
\label{abcd}
\end{eqnarray}
where, in order to make contact with the results in the main text, we have defined
\begin{equation}
\bar \sigma \ = \ 2\,G \, N_c  \sum_f \int \frac{d^4 q}{(2 \pi)^4}  \ g(q^2) \sum_{\lambda=\pm}
\ \hat a^{\lambda,f}_{q_\perp,q_\parallel}\ .
\label{gapapp}
\end{equation}
Given the results in Eq.~(\ref{abcd}), we can easily obtain the expressions
for the functions entering the Ritus transform of the two-point function.
Using Eq.~(\ref{d10}), we get
\begin{equation}
A^{\lambda,f}_{k,p_\parallel} = (1- \delta_{k_\lambda,-1})\, m_c + \bar \sigma \ g^{\lambda,f}_{k,p_\parallel}
\ , \quad
B^{\lambda,f}_{k,p_\parallel} = (1- \delta_{k_\lambda,-1}) \ ,
\quad
C^{\lambda,f}_{k,p_\parallel} = 1\ ,
\quad
D^{\lambda,f}_{k,p_\parallel} = 0\ ,
\label{AD}
\end{equation}
where the definition of $g^{\lambda,f}_{k,p_\parallel}$ is that given in
Eq.~(\ref{mpk}). As we see, $A^{\lambda,f}_{k,p_\parallel}$ coincides with
the expression for $M^{\lambda,f}_{k,p_\parallel}$ given in
Eq.~(\ref{mmain}). Replacing these results in Eq.~(\ref{dapp}) we recover
the expression for $D^{f}_{k,p_\parallel}$ given in Eq.~(\ref{twopoint}). On
the other hand, using the relations in Eqs.~(\ref{d8}) and (\ref{mpk}), we
can write Eq.~(\ref{gapapp}) as
\begin{equation}
\frac{\bar\sigma}{G} \ = \ N_c \sum_{f=u,d} \frac{|q_f B|}{\pi}
\sum_{k=0}^\infty \int \frac{d^2p_\parallel}{(2\pi)^2}
\sum_{\lambda=\pm}\,
\hat A^{\lambda,f}_{k,p_\parallel}\, g^{\lambda,f}_{k,p_\parallel}\ .
\end{equation}
Finally, replacing Eqs.~(\ref{AD}) into Eqs.~(\ref{hatAD}), it is seen that
the expression for $\hat A^{\lambda,f}_{k,p_\parallel}$ coincides with that
given in Eq.~(\ref{alam}). This completes the derivation of the gap
equation, Eq.(\ref{gapmain}), within the framework of the SD formalism
developed, e.g., in Refs.~\cite{Leung:1996qy,Watson:2013ghq,Mueller:2014tea}.

\end{document}